\documentclass[a4paper,11pt]{article}

\usepackage{amssymb, amsmath, lettrine, graphicx, multicol, booktabs, float, subfigure, abstract, xcolor, extarrows, soul}

\usepackage{mathrsfs}

\usepackage{physics}

\usepackage{comment}

\usepackage{mathtools}

\usepackage{jheppub}

\usepackage[english]{babel}
\usepackage[utf8x]{inputenc}
\graphicspath{{figures/}}

\DeclareMathOperator{\e}{e}

\newcommand{\D}{{\rm d}}

\newcommand{\Mpl}{M_{\rm Pl}}
\newcommand{\bfx}{{\vec x}}

\newcommand{\bfk}{{\vec k}}
\newcommand{\bfp}{{\vec p}}

\newcommand{\p}{{\partial}}

\newcommand{\eq}[1]{\begin{equation}#1\end{equation}}

\newcommand{\spl}[1]{\begin{split} #1 \end{split}}
\newcommand{\fg}[1]{\begin{figure}[tbp]\centering #1 \end{figure}}

\newcommand{\Oc}{\mathcal{O}}
\newcommand{\F}{\mathscr{F}}

\newcommand{\kp}{k_{\rm ph}}

\newcommand{\vp}{\varphi}
\newcommand{\s}{\sigma}

\usepackage{cleveref}
\begin{document}
\title{Dissipative Inflation via Scalar Production}

\author[a,b]{Paolo Creminelli,}
\author[c,d]{Soubhik Kumar,}
\author[a,b]{Borna Salehian,}
\author[a,e]{and Luca Santoni}

\affiliation[a]{ICTP, International Centre for Theoretical Physics,
Strada Costiera 11, 34151, Trieste, Italy}
\affiliation[b]{IFPU, Institute for Fundamental Physics of the Universe,
Via Beirut 2, 34014, Trieste, Italy}
\affiliation[c]{Berkeley Center for Theoretical Physics, Department of Physics,
University of California, Berkeley, CA 94720, USA}
\affiliation[d]{Theoretical Physics Group, Lawrence Berkeley National Laboratory, Berkeley, CA 94720, USA}
\affiliation[e]{Universit\'e Paris Cit\'e, CNRS, Astroparticule et Cosmologie, 10 Rue Alice Domon et L\'eonie Duquet, F-75013 Paris, France}
\emailAdd{creminel@ictp.it}
\emailAdd{soubhik@berkeley.edu}
\emailAdd{bsalehia@ictp.it}
\emailAdd{santoni@apc.in2p3.fr}

\abstract{
We describe a new mechanism that gives rise to dissipation during cosmic inflation.
In the simplest implementation, the mechanism requires the presence of a massive scalar field with a softly-broken global $U(1)$ symmetry, along with the inflaton field.
Particle production in this scenario takes place on parametrically sub-horizon scales, at variance with the case of dissipation into gauge fields.
Consequently, the backreaction of the produced particles on the inflationary dynamics can be treated in a \textit{local} manner, allowing us to compute their effects analytically.
We determine the parametric dependence of the power spectrum which deviates from the usual slow-roll expression.
Non-Gaussianities are always sizeable whenever perturbations are generated by the noise induced by dissipation: $f_{\rm NL}^{\rm eq} \gtrsim {O}(10)$.
}
\maketitle
%\tableofcontents
\section{Introduction}
Inflation is the most compelling scenario for the early Universe. However, we are very far from finding its explicit UV realization and its connection to the Standard Model. A more modest goal, which is achievable in the foreseeable future, is to study the {\em qualitative} features of inflation and test them with experiments. The amazing progress in the search for B-modes \cite{BICEP:2021xfz, CMB-S4:2020lpa, LiteBIRD:2023iei} is going to reveal whether inflation occurred at high-energy, with Planckian displacement of the inflaton, or at lower scales \cite{Lyth:1996im}. The possibility that perturbations are generated by a scalar field different from the inflaton will be tested by local non-Gaussianities \cite{Lyth:2002my, Dvali:2003em,Creminelli:2004yq}, once the threshold $f^{\rm loc}_{\rm NL} \sim 1$ is reached (see for example \cite{Dore:2014cca,Castorina:2020zhz,Ferraro:2022cmj}). Equilateral non-Gaussianities will be able to test whether the inflaton perturbations travel subluminally \cite{Cheung:2007st}, although reaching $f^{\rm eq}_{\rm NL} \sim 1$ will require more patience (see for example \cite{Castorina:2020zhz,Cabass:2022epm}). 

Another qualitative feature that one would like to test is whether the time-dependent inflaton background dissipates energy, producing additional degrees of freedom. One possibility is that the produced degrees of freedom thermalize after the production: this is the case of warm inflation, in which the inflaton coexists with a thermal plasma. This scenario has a long history \cite{Berera:1995ie}, but it found only recently a compelling, minimal realization \cite{Berghaus:2019whh}. Cosmological perturbations in this case are generated by the coupled evolution of the inflaton and the thermal bath. The model gives a rich phenomenology of non-Gaussianity, detectable in a wide range of parameter space \cite{Mirbabayi:2022cbt}. Another possibility is that the produced degrees of freedom do not thermalize and this is the scenario we are going to explore in this paper.

A simple mechanism for dissipation is given by the pseudoscalar coupling of the inflaton $\phi$ with gauge fields: $\phi F \tilde F$. This coupling is quite natural since it is compatible with the approximate shift symmetry of the inflaton action. As explained by Anber and Sorbo \cite{Anber:2009ua}, around the time-dependent inflaton background the gauge fields develop an instabilility and are copiously produced. This process backreacts on the inflaton, generating friction, which adds to the usual Hubble friction. Cosmological perturbations do not arise from the usual vacuum fluctuations but are given by fluctuations in gauge-field production.

The model we are going to study below can be seen as a {\em scalar} version of the gauge-field scenario. A complex scalar $\chi$ develops a window of instability because of the interplay between the mass term and a coupling  $i {\partial_\mu\phi}\left(\chi\partial^\mu\chi^* -\chi^*\partial^\mu\chi\right)$, which plays a role analogous to the $\phi F\tilde{F}$ coupling described above.\footnote{A similar coupling appears also in the context of fermion production in axion inflation, where $\partial_\mu\phi$ couples to the fermionic current $J^\mu_A = i \bar \psi \gamma^\mu \gamma^5 \psi$ of a (softly broken) axial $U(1)_A$ symmetry~\cite{Adshead:2015kza,Adshead:2018oaa}.}  As always, there is more model-building flexibility in dealing with scalars compared to gauge fields and this will allow us to address some important drawbacks of the Anber--Sorbo model (at the price of less minimality).
 
The main qualitative difference between our scenario and the case of gauge-field instability is that the window of instability in our case will be on scales much shorter than Hubble, while in the vector case the instability persists until Hubble crossing. (The dynamics of $\chi$ will be studied in~\cref{sec:model}, and appendices~\ref{app:canquant} and \ref{app:wkb}.) This difference has two important consequences. First, in a single Hubble patch, there are many short modes produced by the instability. The average of many modes suppresses the power spectrum (see~\cref{sec:perturbations}), while in the original Anber--Sorbo model one has to introduce a huge number of distinct gauge bosons in order to get an analogous suppression. The large number also makes the statistics of perturbation close to Gaussian, by the central limit theorem, and thus compatible with observations (see also \cite{Adshead:2015kza,Adshead:2018oaa}). The second implication of producing modes on short scales is that their response is much faster than Hubble and can be treated as instantaneous. On the contrary, the typical timescale for the backreaction of the gauge fields is the Hubble time: this introduces nonlocality in the equations, and more importantly, gives rise to large oscillations of the background solution. The solution ceases to be an attractor as recently understood both using numerical simulations and analytical arguments; see \cite{Peloso:2022ovc} and references therein. 

In our model, the effect of the produced degrees of freedom can be treated as a modification of the inflaton Lagrangian that is local in time (plus a noise term). From this point of view, the model can be seen as an explicit realization of the general Effective Field Theory (EFT) of dissipative inflation studied in \cite{LopezNacir:2011kk}. The advantage of our setup is that the production of the extra degrees of freedom is compatible with the inflaton shift symmetry. In previous dissipative models, like trapped inflation \cite{Green:2009ds}, the production of particles is associated with the breaking of the shift symmetry acting on the inflaton: in this case one needs some additional model-building gymnastic to control the flatness of the inflaton potential.

As for the other features discussed at the beginning, non-Gaussianity is the key experimental signal of dissipation. There are actually two sources of non-Gaussianity (see section~\ref{sec:nongaussianity} and appendix~\ref{app:2pt}): one originating from the inflaton nonlinearities and the second from the statistics of the noise. The first category has a contribution that depends on the amount of dissipation: if the linear equation of motion contains a modified friction $(3 H + \gamma) \dot\varphi$, then the nonlinear realization of Lorentz invariance implies a non-Gaussian contribution with a size $f^{\rm eq}_{\rm NL} \sim \gamma/H$ \cite{LopezNacir:2011kk}. (This is similar to what happens when considering models where the inflaton has a reduced speed of propagation.) Therefore reaching the threshold $f^{\rm eq}_{\rm NL} \sim 1$ will also test whether the inflaton perturbations have a sizeable dissipation. In the model that we study in this paper, we also find that the interaction of the inflaton with the noise generates $f^{\rm eq}_{\rm NL}\gtrsim O(10)$, approximately independent of $\gamma/H$, which might be within the reach of upcoming cosmological surveys.
On the other hand, the non-Gaussianity of the noise, which is also model-dependent and cannot be estimated with EFT considerations, turns out to be subdominant in our setup.

\paragraph{Notation and Conventions.}   We work in mostly plus metric signature $(-, +, +, +)$, and natural units $c=\hbar=1$. 
We denote the reduced Planck mass by $\Mpl=(8\pi G)^{-1/2}$.
 We adopt the Fourier convention
\begin{equation}
 f(t,\bfx)=
 \int\frac{\D^3\bfk}{(2\pi)^{3}}\,
 \e^{i\bfk\cdot\bfx}f(t,\bfk)\,.
\end{equation}
We will sometimes use the following shorthand: $\int_\bfk \equiv \int\frac{\D^3\bfk}{(2\pi)^{3}}$. 
We will denote with $\tau$ the conformal time, defined by $\D\tau= \D t/a(t)$, where $a$ is the scale factor.

\section{Overview of the Mechanism and Background Evolution}
\label{sec:model}
We consider a complex scalar field $\chi$ which is charged under a (softly broken) global $U(1)$ symmetry.
Since $\chi$ is not the inflaton candidate in our scenario, we will often refer to $\chi$ as the  Additional Degree Of Freedom (ADOF).
The action describing the coupling between the inflaton $\phi$ and the ADOF $\chi$, in the presence of gravity, is given by
\begin{multline}
S = \int \D^4x \sqrt{-g} \, \bigg[ \frac{1}{2}\Mpl^2 R -\frac{1}{2}(\partial\phi)^2 - V(\phi) - |\partial\chi|^2 + M^2|\chi|^2
\\
		 - i\frac{\partial_\mu\phi}{f}\left(\chi\partial^\mu\chi^* -\chi^*\partial^\mu\chi\right)-\frac{1}{2}m^2 (\chi^2+\chi^*{}^2) \bigg] \, .
\label{action}
\end{multline}
This model was also considered in~\cite{Bodas:2020yho}, in the context of cosmological collider physics~\cite{Chen:2009zp, Arkani-Hamed:2015bza}, in the limit of small backreaction from the produced $\chi$ quanta.
In the above, $M$, $m$, and $f$ have mass dimensions $+1$. 
We have also chosen an unusual sign of the mass term $M^2|\chi|^2$ for later convenience. 
We will confirm below that this choice is safe in the sense that it does not induce any tachyonic instability for the superhorizon modes.
In the absence of the last term in \cref{action}, the action is exactly invariant under the $U(1)$  transformation $\chi\rightarrow\e^{i \alpha}\chi$, with $\alpha\in \mathbb{R}$. 
Thanks to this enhanced symmetry, the hierarchy $m^2\ll M^2$---which we assume below---is protected against large radiative corrections.
The inflaton potential $V(\phi)$ is the only source of breaking of the shift symmetry  $\phi\rightarrow\phi + {\rm constant}$, which becomes exact in the limit in which $V$ is $\phi$-independent. 
We will also assume that $V(\phi)$ is even under $\phi\rightarrow-\phi$, in such a way that the action \eqref{action} enjoys the following exact symmetry: $\phi\rightarrow-\phi$ and $\chi\rightarrow\chi^*$.\footnote{This  will be relevant in \cref{sec:computationexpvalues} (see, in particular, footnote~\ref{foot:sign}) and \cref{app:signofO}.} 

To simplify the presentation, we have kept operators only up to dimension-5 in~\cref{action}. As we will see later, we must assume that $m^2$ and $M^2$ are functions of $(\p\phi)^2$---which preserves the $\phi$ shift symmetry---and thereby include certain dimension-6 and higher contributions, while still demanding a controlled EFT. In addition, for simplicity, we will only consider quadratic terms in $\chi$. 

It is useful to formulate our model also in terms of the EFT of inflation \cite{Cheung:2007st,Senatore:2010wk}. The $\chi$ sector is described in unitary gauge by
	\begin{multline}
		S = \int \D^4x \sqrt{-g} \, \bigg[ - |\partial\chi|^2 + M^2 |\chi|^2
		- i\rho\left(\chi\partial^0\chi^* -\chi^*\partial^0\chi\right)-\frac{1}{2}m^2 (\chi^2+\chi^*{}^2) \bigg] \, .
		\label{actionEFTI}
	\end{multline}
	where $\dot\phi_0(t)\equiv \rho f$. The masses $M^2$ and $m^2$ will in general depend on $g^{00}$ (we are going to need only the unperturbed value and the first perturbation, proportional to $(g^{00}+1)$). The only new ingredient is the presence of a 1-derivative operator for $\chi$. This operator is possible only in the presence of two (real) scalars, otherwise it reduces to a total derivative, and opens up the possibility of a window of instability for $\chi$.

\subsection{Homogeneous Solution and Particle Production}
In this section, we study the equation of motion for $\chi$ on the unperturbed, homogeneous, and isotropic inflationary background. 
We will denote the time-dependent background profile for $\phi$ as $\phi_0(t)$, and work at the leading order in the slow-roll approximation, i.e.,~$|\dot{H}|\ll H^2$ and $|\ddot{\phi}_0| \ll H |\dot{\phi}_0|$.
Without loss of generality, we will assume that $\dot{\phi}_0>0$.
As we will see, for certain choices of the parameters in the model~\eqref{action}, the motion of the inflaton $\phi$ induces instabilities in the ADOF sector and leads to a significant production of $\chi$ particles.
We discuss the slow-roll background evolution and backreaction effects in more detail in section~\ref{sec:slow-roll} below.

We start with the $\chi$  equation of motion, obtained from the action \eqref{action}:
\begin{equation}
\label{eq:chi1}
\square\chi + \frac{2i}{f}\nabla^\mu\phi\nabla_\mu\chi + \left(M^2+i\frac{\square\phi}{f}\right)\chi - m^2\chi^* = 0 \, .
\end{equation}
We also write $\chi$ in terms of two real fields $\chi_1$ and $\chi_2$,
\begin{equation}
\chi \equiv \frac{\chi_1 + i\chi_2}{\sqrt{2}} \, ,
\end{equation}
and set $\phi_0(t)\equiv \rho f t$, where $\rho$ has mass dimension $+1$. 
Then, rescaling the field components $\chi_1$ and $\chi_2$  as $\chi_i \equiv  \sigma_i/a^{3/2}$, for $i=1,2$,  the ADOF equation of motion \eqref{eq:chi1} on the homogeneous background $\phi_0(t)$ takes the form:
\begin{subequations}
\label{eqschi1chi2dS2}
 \begin{align}
 	\ddot{\sigma}_1 -  \frac{\vec{\nabla}^2\sigma_1}{a^2}  - \left(M^2-m^2+\frac{9H^2}{4}\right)\sigma_1 -2\rho\dot\sigma_2 & = 0\, ,\\
	\ddot{\sigma}_2 -  \frac{\vec{\nabla}^2\sigma_2}{a^2}  - \left(M^2+m^2+\frac{9H^2}{4}\right)\sigma_2 +2\rho\dot\sigma_1 & = 0\, .
 \end{align}
\end{subequations}
We can decompose $\sigma_1$ and $\sigma_2$  in terms of creation and annihilation operators as
\begin{equation}
	\s_i(t,\bfx) = \int \frac{\D^3\bfk}{(2\pi)^{3}} \e^{i\bfk\cdot\bfx} \left[
	(F_k(t))_{ij}\hat{a}_j(\bfk)  + (F_k^*(t))_{ij}\hat{a}^\dagger_j(-\bfk)  
	\right] \, ,
\label{quantsigma}
\end{equation}
where  $F_k(t)$ is a $2\times2$ matrix, whose elements are functions of $k$ and $t$, and $\hat{a}(\bfk)$ is a 2-vector of annihilation operators (see appendix~\ref{app:canquant} for details). 
To solve for the mode function matrix $F_k(t)$ on the inflationary background we will use the standard WKB approximation. 
This assumes that the coefficients in \eqref{eqschi1chi2dS2} depend so weakly on time that all the time dependence in  $F_k(t)$ is essentially encoded in a common phase factor $\e^{i \int \omega_{\pm}\D t}$. 
The (time-dependent) WKB frequencies $\omega_{\pm}$ are given by
\eq{
\omega_{\pm}^2=\left(\sqrt{\frac{k^2}{a^2}+\mu^2}\pm\rho\right)^2-\frac{m^4}{4\rho^2}\,,
\label{omegapm}
}
where we defined 
\eq{
	\mu^2\equiv\rho^2-M^2-\frac{9H^2}{4}+\frac{m^4}{4\rho^2}\,.
\label{nu2}
}
\fg{  
\includegraphics[width=0.6\textwidth,trim={0 2.8cm 0 1.2cm},clip]{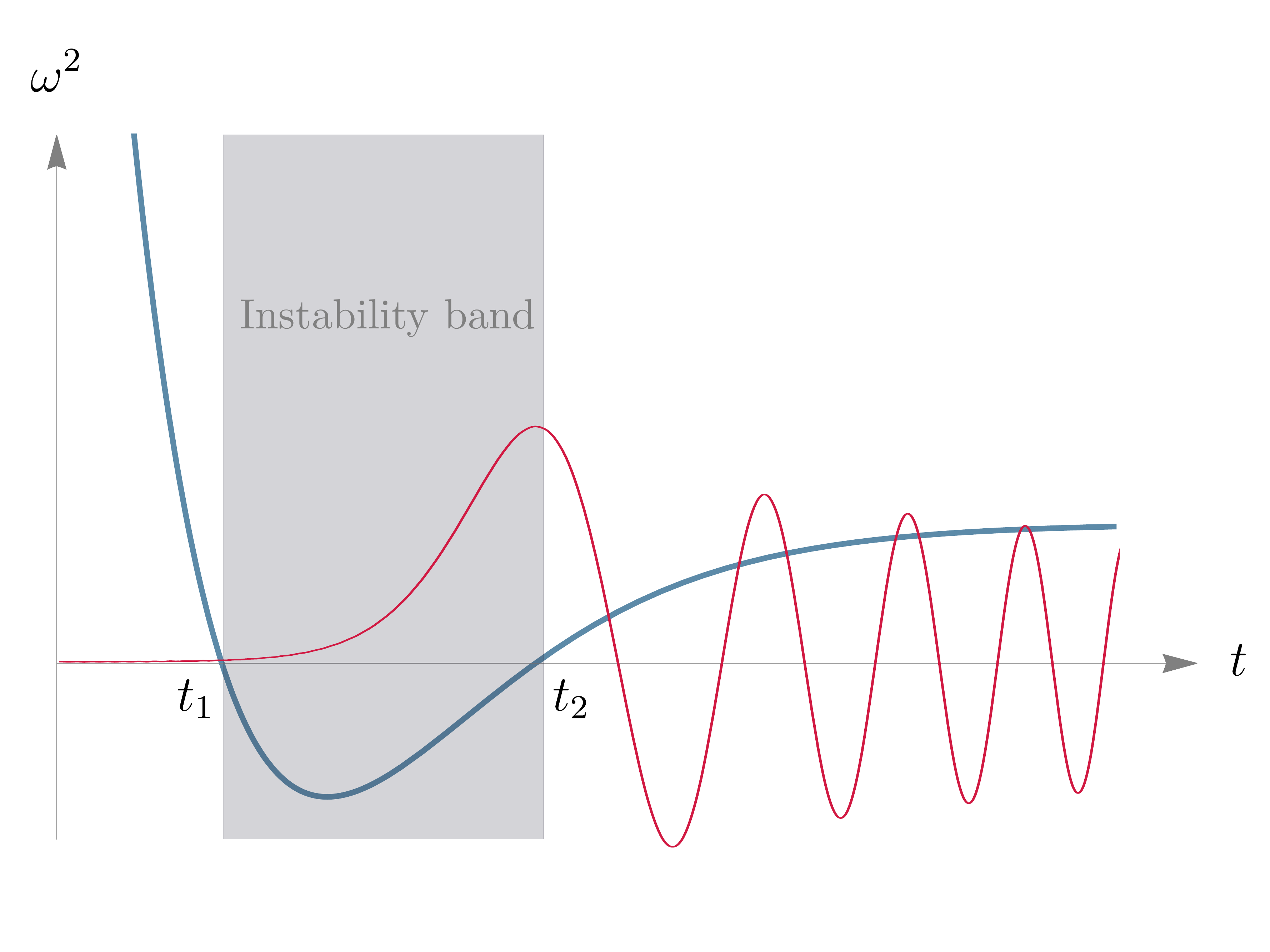}
\caption{The blue thick line is the plot of $\omega_-^2$ as a function of time (in arbitrary units) for fixed $k$ and given choice of the mass parameters in \eqref{omegapm}.  The thin red line is instead a sketch of the profile of the corresponding mode function. Note that, at very early times, $\omega_-^2>0$ and the mode behaves as a plane wave satisfying the standard Bunch--Davies initial condition (the amplitude is too small to be visible in the plot). After entering the instability band ($t_1<t<t_2$), $\omega_-^2$ turns negative and the mode starts growing exponentially. For $t>t_2$, the mode oscillates again, but with a much larger amplitude. Its relative enhancement compared to the early-time wave amplitude can be read off from the WKB solution \eqref{fullWKB-main}.}
\label{omega2plot}
}%
\noindent Corrections to the leading WKB solution can be systematically accounted for by solving for the time dependence of the amplitude of the mode functions. 
We review in detail the derivation of the WKB solution (up to the desired order) in appendix~\ref{app:wkb}. Here, we just report the result and comment on the aspects that will be useful in the following.
First we note that, provided 
\begin{equation}
	m^2<M^2+\frac{9H^2}{4}<\rho^2+\frac{m^4}{4\rho^2}<2\rho^2\,,
	\label{eneq}
\end{equation}
the mode function corresponding to $\omega_-$ develops an instability, i.e., $\omega_-^2<0$, when the physical momentum $k/a$ takes values  in the finite range (see figure~\ref{omega2plot})
\begin{align}
	-m^2 + M^2 + \frac{9}{4}H^2 < \frac{k^2}{a^2} < m^2 + M^2 + \frac{9}{4}H^2 \, .
\label{ineqka}
\end{align}
This does not happen for the $\omega_+$ mode which satisfies $\omega_+^2>0$ at all times. 
We can write the full WKB solution by first identifying the points in time where $\omega_-^2=0$ for a fixed momentum $k$.
We denote these two points by $t_1$ and $t_2$ ($t_1<t_2$).
The WKB solution can thus be obtained by solving the equation for $F_k(t)$ away from $t_1$ and $t_2$, such that the solution is well defined and not singular, and by then performing the matching across $t_1$ and $t_2$. 
The final result can be summarized as follows: 
\begin{align}
	\label{fullWKB-main}
	F_k (t)  &=
	\begin{cases}
	\bigg(\vec{Q}_+\,\,,\,\, i\vec{Q}_+\bigg)\e^{-i\int^t\omega_+ }+\bigg(\vec{Q}_-\,\,,\,\,-i\vec{Q}_-\bigg)\e^{-i\int^t\omega_-} 
		   ,      &  t<t_1 \\ \\
	\e^{-i\theta_1} \bigg(\vec{Q}_-\,\,,\,\,-i\vec{Q}_-\bigg)\e^{\int_{t_1}^t|\omega_-|}
		 ,  % + \text{ ``$\omega_+$ mode"}  ,    
		  & \hspace{-0.7cm}  t_1< t<t_2 \\\\
		 \e^{-i\theta_1+\pi\xi}  \left[
		\bigg(\vec{Q}_-\,\,,\,\,-i\vec{Q}_-\bigg)\e^{-i\int_{t_2}^t\omega_-}+i \bigg(\vec{Q}^*_-\,\,,\,\,-i\vec{Q}^*_-\bigg)\e^{+i\int_{t_2}^t\omega_-}  \right] , %+ \text{``$\omega_+$ mode"}
				&  t>t_2
	\end{cases}
\end{align}
where we introduced the phase $\theta_1=\int^{t_1}\dd{t'}\omega_{-}(t')$. The 2-vectors $\vec{Q}_\pm$, denoting the columns of the matrices in \eqref{fullWKB-main}, and the growth exponent $\xi$  can be read off from   eqs.~\eqref{Qpmwkb-app}  and \eqref{xi-app}, respectively. 
Note that in the second and third line of \eqref{fullWKB-main} we omitted to write explicitly the terms involving the $\omega_+$ modes, whose amplitudes are exponentially suppressed compared to the $\omega_-$ ones, and are therefore negligible.
Since the $\omega_+$ modes will play no role in the following discussion, we will neglect them altogether, and from now on drop the minus subscript on the growing modes.
Those frequencies will be denoted simply by $\omega$ instead of $\omega_-$. 
We will reinstate the full notation when explicitly needed.

%==========================================================================================

\subsection{Local Response and Hierarchies}
\label{sec:localapprox}
The  instability in the $\chi$ spectrum  occurs whenever the conditions \eqref{eneq} are satisfied. In the following, to simplify the analysis, we will consider physical situations in which the particle production happens on scales much shorter than the Hubble size. 
In particular, we will require the time scale for dissipation induced by the ADOF to be much smaller than $H^{-1}$. 
This is a necessary condition in order to be able to describe the dissipative dynamics of the system in a local approximation and to neglect memory effects. 

Working in a local regime has multiple advantages.  
First, a delayed response to a change in the inflaton velocity, which is generically expected whenever the typical time scale of particle production becomes comparable to $H^{-1}$, is absent in a local regime.
This can avoid the possible presence of additional resonances and instabilities. 
These latter effects have been, for instance, explicitly seen in the context of axion inflation~\cite{Domcke:2020zez,Caravano:2022epk,Peloso:2022ovc}, where oscillatory features have been found in the inflaton velocity and the gauge-field spectrum, as a result of a nonlocal response. 
On the other hand, locality will allow us to obtain an analytic description of the dynamics of the perturbations~\cite{LopezNacir:2011kk}. 
In addition, in the strong-backreaction regime, the local approximation provides a natural way of suppressing potentially large contributions in the induced scalar correlators, without the need of a large number of ADOFs~\cite{Anber:2009ua}, as we will explicitly see in section~\ref{sec:perturbations}.

To this end, we  require that the $\chi$ modes enter the instability phase when they are still subhorizon, i.e.,
\begin{equation}
 \frac{k}{a} \sim M \gg H \, .
\label{instwindowc}
\end{equation}
In addition, a necessary condition for locality is that the instability band is narrow in Hubble units, i.e.,
\begin{equation}
H(t_2-t_1)\sim\frac{\Delta(k/a)}{k/a}  \sim \frac{m^2}{M^2} \ll 1 \, ,
	\label{instwindownarrow}
\end{equation}   
where we have used $\Delta(k/a) \sim m^2/M$.
We are interested in cases with significant particle production. Therefore, we require the parameter $\xi$ in the exponential factor $\e^{\pi\xi}$ in \cref{fullWKB-main} to be $\xi \gtrsim O(1)$. Using \cref{instwindownarrow} in the expression for $\xi$ in \cref{xi-app}, we obtain: 
\begin{equation}
\xi \simeq \frac{m^4}{8H \rho M^2}\gtrsim O(1)\,.
\label{xiapp}
\end{equation} 
Note that, from eqs.~\eqref{eneq}, \eqref{instwindowc}, and \eqref{instwindownarrow}, it follows that $m\ll \rho$ and $M\lesssim\rho$.
Thus, the inequality \eqref{xiapp} can be satisfied only if we assume that $H\ll m$.

As we will explicitly see later on, the equation for the inflaton perturbations is modified by particle production. One effect that is induced is an additional friction term. The requirement that such extra friction coefficient is moderately, but not exponentially, large implies that (see \cref{gamma})\footnote{{More precisely, using~\cref{gamma} we find $\gamma/H \sim \xi^2 \exp(2 \pi \xi)M^2/f^2$. The requirement of $\gamma\gtrsim H$, but with $\gamma$ not larger than $10^2H$ (see \cref{fNLgammaH}), gives the relation in~\cref{expf}.}}
\eq{
\e^{2\pi\xi}\frac{M^2}{f^2} \lesssim1 \quad \implies \quad f\gg M\,.
\label{expf}
}
The previous conditions are schematically summarized by the following hierarchy:
\begin{equation}
f\gg \rho \gtrsim  M \gg m \gg H \, .
\label{hierarchy}
\end{equation}
Although necessary, we stress that \eqref{hierarchy} is not sufficient to completely ensure the absence of nonlocal effects in our setup.
Couplings of the inflaton to conserved quantities or operators with constant averages over long time scales can generically still induce long-distance effects and delayed response. 
We will come back to this point with a concrete example when we discuss more about locality in section~\ref{sec:perturbations}.

Without loss of generality, for the sake of the presentation, we will assume from now on that the scales $\rho$ and $M$ are of the same order, and use them interchangeably in the estimates below, dropping irrelevant order-one factors.

\subsection{Backreaction Effects and Slow-roll Background Evolution}
\label{sec:slow-roll}
The exponentially amplified $\chi$ modes will eventually backreact on the inflationary evolution. 
At the background level,  two main types of backreaction effects can be distinguished. 
First, the large production of $\chi$ particles extracts energy from the inflaton, providing a new source of dissipation that can potentially overcome the Hubble friction in the $\phi_0$ equation of motion. 
In addition, the energy density of the produced particles contributes to the  Friedmann equations. 
In this work, we will mainly focus on the large-backreaction regime in which the dissipation due to $\chi$ production becomes comparable to, or larger than, the standard Hubble friction. 
In such a regime, the inflationary evolution is dominated, in large part, by dissipation into the ADOF sector, deviating from the standard slow-roll scenario at least by an order-one amount.
On the other hand, we will require the energy density of the $\chi$ modes in the Friedmann equations to be subleading compared to the inflaton energy density, which remains dominated by the inflaton potential. 

To be as general as possible, we will allow $M^2$ and $m^2$  to depend on $\phi$ through the shift-symmetric combination $X\equiv-(\p\phi)^2/(2\rho^2 f^2)$.\footnote{In fact, as we will explicitly discuss in \cref{sec:attractor}, having $M^2$ and $m^2$ nontrivial   functions of $X$ is not just for the sake of generality, but is  a necessary ingredient for the background solution to be an attractor.}  
The action \eqref{action} should thus be regarded as an effective theory with a cutoff scale $\Lambda \simeq ( \rho f)^{1/2}$. In particular, we will expand the functions $M^2(X)$ and $m^2(X)$ in derivatives as
\begin{equation}
M^2(X)= M_0^2 \left(c_0 +  c_1 \frac{(\p\phi)^2}{\rho^2f^2} + \ldots \right) \, , \qquad m^2(X)= m_0^2  \left(c'_0 +  c'_1 \frac{(\p\phi)^2}{\rho^2f^2} + \ldots \right) \, ,
\label{M2m2expansion}
\end{equation}
where $c_n$ and $c'_n$ are order-one  couplings, while $M_0\sim \rho$ and $m_0\ll\rho $, as in the hierarchy \eqref{hierarchy}.\footnote{A careful reader might have noticed that the first operator in the second line of \cref{action} enters at a scale $f$ that is parametrically larger than $\Lambda$, associated with the $X^n$ terms in the expansions \eqref{M2m2expansion}. We stress that this is a consistent choice, since such an operator is the only one that breaks the $Z_2$ symmetry $\phi\rightarrow-\phi$ and cannot, therefore, be generated quantum mechanically from the other $Z_2$-symmetric operators in the action.} 
Note that, on the background solution $\phi_0(t)= \rho f t$, the $X^n$ terms with $n>1$ in the expansions \eqref{M2m2expansion} provide at most order-one corrections to the leading $n=0$ terms, i.e.,~$M\sim M_0$ and $m\sim m_0$.

From the action  \eqref{action}, we can now compute the  equations of motion for $\phi$, which after using \eqref{eq:chi1}, can be cast in the form
\begin{equation}
\nabla_\mu\left[\left(1+\frac{|\chi|^2}{\rho^2f^2}(M^2_X-2\rho^2)-\frac{1}{2\rho^2f^2}(\chi^2+\chi^*{}^2)m^2_X\right)\nabla^\mu\phi\right] -V'(\phi) +\frac{im^2}{f}(\chi^2-\chi^*{}^2) = 0 \, ,
\label{phi-eq}
\end{equation}
where we introduced the shorthand notation $M_X^2\equiv \frac{\D M^2}{\D X}$ and $m_X^2\equiv \frac{\D m^2}{\D X}$.
The energy-momentum tensor for the matter fields is
\begin{multline}
T_{\mu\nu}=\left[1+\frac{|\chi|^2}{\rho^2f^2} M^2_X-\frac{1}{2\rho^2f^2}(\chi^2+\chi^*{}^2)m^2_X\right]\p_\mu\phi\p_\nu\phi+2\p_{(\mu}\chi\p_{\nu)}\chi^*
\\
+\frac{2i}{f}\p_{(\mu}\phi\left( \chi\p_{\nu)}\chi^*-\chi^*\p_{\nu)}\chi\right)+g_{\mu\nu}\mathcal{L} \, ,
\label{Tmunu}
\end{multline}
where $\mathcal{L}$ is the Lagrangian density in \eqref{action}.

Backreaction effects induced by the ADOF can be accounted for by taking the expectation values of the $\chi$ operators in  the  Friedmann equations. Using \eqref{Tmunu}, where we evaluate $\phi$ and the metric on the background, we find the Friedmann equations
\begin{multline}
3\Mpl^2H^2   = \frac{\dot{\phi}_0^2}{2} +  V + \ev{|\dot{\chi}|^2} +\frac{1}{a^2}\ev{|\p_i\chi|^2} 
- \left( M^2-X M^2_X \right)\ev{\vert\chi\vert^2} 
\\
 + i\rho\ev{\chi \dot{\chi}^*-\dot{\chi}\chi^*}
+  \frac{1}{2}\left( m^2-X m^2_X \right)\ev{\chi^2 + \chi^{*2}},
\label{Friedmann1}
\end{multline}
\begin{equation}
-\Mpl^2\dot H   = \frac{\dot{\phi}^2_0}{2}+\ev{|\dot{\chi}|^2}+\frac{1}{3a^2}\ev{|\p_i\chi|^2} 
+X M^2_X\ev{\vert\chi\vert^2} 
+ i\rho\ev{\chi \dot{\chi}^*-\dot{\chi}\chi^*}
-  \frac{1}{2}X m^2_X  \ev{ \chi^2 + \chi^{*2} },
\label{Friedmann2}
\end{equation} 
while the background equation  for the  inflaton  takes the form
\begin{equation}
\frac{1}{a^3}\partial_t \left[\left(1+\frac{\ev{|\chi|^2}}{\rho^2f^2}(M^2_X-2\rho^2)-\frac{1}{2\rho^2f^2}\ev{\chi^2+\chi^*{}^2}m^2_X\right)a^3\dot \phi_0\right] + V'(\phi) - \frac{im^2}{f}\ev{\chi^2-\chi^*{}^2} = 0 \, .
\label{scalareqback}
\end{equation}
We stress that, in eqs.~\eqref{Friedmann1}, \eqref{Friedmann2}, and \eqref{scalareqback}, the quantities $X$, $M^2$, $m^2$ and derivatives thereof are implicitly all meant to be computed on $\phi_0(t)$.

\subsubsection{Computation of the Expectation Values and Slow-roll Conditions}
\label{sec:computationexpvalues}
We will now compute the average of the operators appearing in the background equations \eqref{Friedmann1}--\eqref{scalareqback}. 
We will see that it is possible to identify two different types of operators, depending on whether they have nontrivial support only on sub-Hubble scales or not.

Let us start by estimating the expectation value $\ev{\vert\chi\vert^2} $. Using the definition $\chi \equiv  \sigma/a^{3/2}$ and eq.~\eqref{quantsigma}, we can write
\begin{equation}
\ev{\vert\chi\vert^2} = \frac{1}{2} \ev{\chi_1^2+\chi_2^2} 
= \frac{1}{4\pi^2a^3}\int \D k \, k^2 \left( [F_k(t)\cdot F_k^\dagger(t)]_{11} + [F_k(t)\cdot F_k^\dagger(t)]_{22}   \right) \, .
\label{modchi2-1}
\end{equation}
The average \eqref{modchi2-1} is a sum over all $\chi$ modes. To get an estimate, we shall neglect the integral over the modes that, with sufficiently large $k\equiv \vert\bfk\vert$, have not entered the instability phase at a given time $t$ yet, and we shall focus instead only on those whose amplitudes are exponentially enhanced. 
This provides a `UV cutoff', set by the scale $k/a\sim M$, up to which we shall compute the integral~\eqref{modchi2-1}. 
Using  \eqref{fullWKB-main}  for $t>t_1$ and \eqref{Qpmwkb-app}, the leading contribution to \cref{modchi2-1} is, by simple dimensional analysis,
\begin{equation}
\ev{\vert\chi\vert^2} \simeq \frac{\e^{2\pi\xi}}{4\pi^2a^3}\int \D k \, k^2  \vert Q_-\vert^2   \simeq  \frac{\e^{2\pi\xi}}{4\pi^2}M^2 \, ,
\label{modchi2-2}
\end{equation}
where we neglected oscillatory terms and used the fact that the integral gets contributions from all $k/a$'s from small values  up to the scale $M$ (see figure~\ref{vevO}).\footnote{Note that the integrand in \eqref{modchi2-2} has a nonzero (almost) constant average on Hubble scales and  can  therefore induce a nonlocal response at the level of the perturbations. In other words, a change in the inflaton velocity can affect $\ev{\vert\chi\vert^2}$ at much later times. We will discuss in section~\ref{sec:perturbations} under what conditions these nonlocal effects can be absent.}

Next, we look at the operator $\chi^2+\chi^*{}^2$. Its expectation value can be written as
\eq{
\ev{\chi^2+\chi^*{}^2}=\ev{\chi_1^2-\chi_2^2}=\frac{1}{2\pi^2a^3}\int \D k \, k^2 \left( [F_k(t)\cdot F_k^\dagger(t)]_{11} - [F_k(t)\cdot F_k^\dagger(t)]_{22}   \right) \, .
}
We note that $\ev{\chi^2+\chi^*{}^2}$ is subdominant compared to $\langle\vert\chi\vert^2\rangle$ since the former vanishes in the limit of $m\rightarrow 0$.\footnote{One can check that for $m=0$ it is possible to decouple the two fields by means of a suitable field redefinition which implies $\ev{\chi_1^2}=\ev{\chi_2^2}$.} In particular, following the same steps as in \eqref{modchi2-2}, we obtain
\begin{equation}
\ev{\chi^2+\chi^*{}^2} \simeq \frac{\e^{2\pi\xi}}{4\pi^2}m^2 \, .
\label{chipluschi}
\end{equation}
 Furthermore, taking into account the different dimensionful prefactors multiplying the two operators in eqs.~\eqref{Friedmann1}--\eqref{scalareqback}, it follows that the terms proportional to $\ev{\chi^2+\chi^*{}^2}$ in \eqref{Friedmann1}--\eqref{scalareqback} are even more suppressed compared to $\ev{\vert\chi\vert^2}$:
\begin{equation}
\frac{m^2_X \ev{\chi^2+\chi^*{}^2} }{ (M_X^2-2\rho^2) \ev{\vert\chi\vert^2} } \simeq   \frac{m^4}{\rho^4}  \ll 1 \, ,
\label{ratiochi2chi2}
\end{equation}
where we have assumed $M_X^2-2\rho^2\sim M^2$ and $m_X^2\sim m^2$. 

\fg{
	\includegraphics[width=0.45\textwidth]{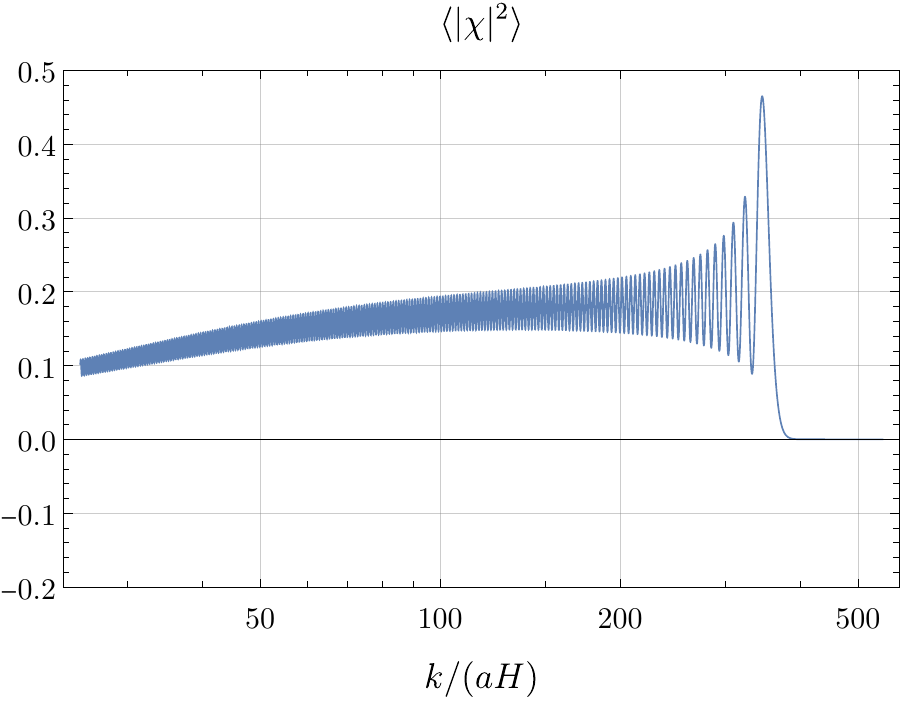}\hfill
	\includegraphics[width=0.45\textwidth]{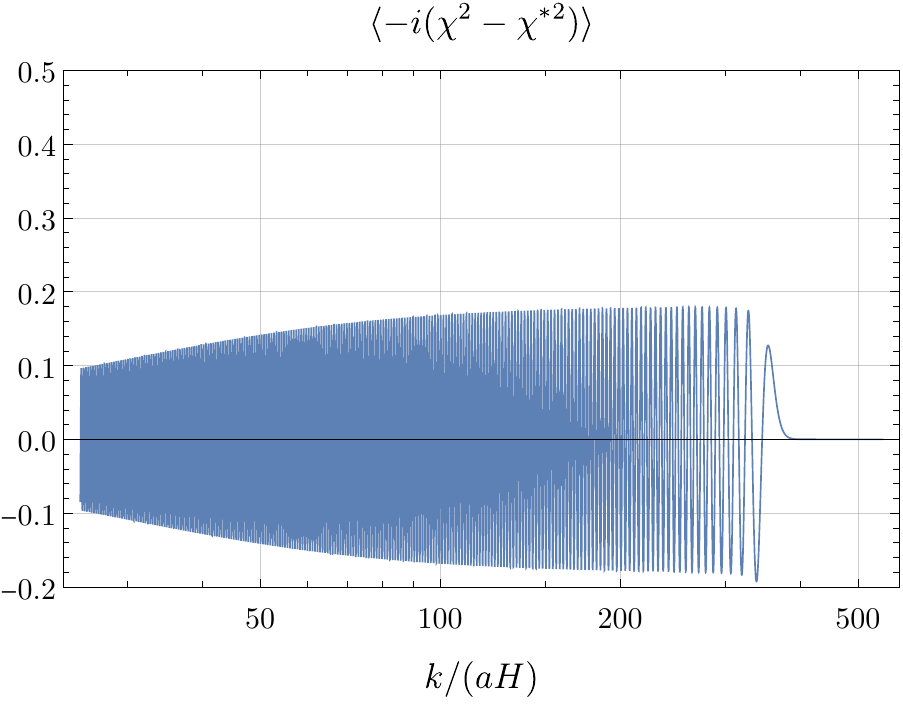}
	\caption{Numerical plots of the integrands in eqs.~\eqref{modchi2-1} (left) and \eqref{eqschiphibis-last} (right),  as functions of $k/(aH)$, for the computation of the  expectation values of $\ev{|\chi|^2}$ and $\ev{-i(\chi^2-\chi^*{}^2)}$, respectively. The vertical scale is arbitrary. For the plots we have set the instability scale at $k/(aH)\sim400$. We can distinctly distinguish two  types of behavior:  the function oscillates either around zero (right) or around a nonzero value (left). This implies that, when we compute the integral in \cref{modchi2-1} to get $\ev{|\chi|^2}$, all modes with $k/(aH)\leq 400$ will contribute. Instead, only the modes localized around the instability window will be relevant for the integral  \eqref{eqschiphibis-last} for $\ev{-i(\chi^2-\chi^*{}^2)}$, while all the others will average out. }
	\label{vevO}
}

Following the same logic, it is instructive to consider the operator $(\chi^2-\chi^{*2})$ appearing in \eqref{scalareqback}, and compare its expectation value with \eqref{modchi2-2} and \eqref{chipluschi}. We  have:
\begin{equation}
\label{eqschiphibis-last}
 -i\ev{\chi^2-\chi^{*2}} = 2 \ev{\chi_1\chi_2}
 =
\frac{1}{2\pi^2a^3}\int \D k \, k^2 [F_k(t)\cdot F_k^\dagger(t)]_{12} \, .
\end{equation}
As opposed to $\ev{\vert\chi\vert^2}$ and $\ev{\chi^2+\chi^*{}^2}$, the integrand in   \eqref{eqschiphibis-last} oscillates around a zero mean value as a function of $k/a$, except for the finite range of $k/a$ where the modes are exponentially enhanced. 
In other words, the leading contribution to the integral \eqref{eqschiphibis-last} has support over the instability window only (see figure~\ref{vevO}). We can thus estimate: 
\begin{equation}
\label{eqschiphibis-last2}
 -i\ev{\chi^2-\chi^{*2}}  
 \simeq 
\frac{\e^{2\pi\xi}}{2\pi^2}m^2 \, ,
\end{equation}
where we used eqs.~\eqref{instwindowc} and \eqref{instwindownarrow}, and set $\rho \simeq M$. We provide additional details on the derivation of \eqref{eqschiphibis-last2} in appendix~\ref{app:signofO}, where  we show in particular that assuming $\dot{\phi_0}>0$ the right-hand side of \eqref{eqschiphibis-last2} has positive sign, i.e.,
\begin{equation}
-i\ev{\chi^2-\chi^{*2}}  > 0 \, ,
\label{posO}
\end{equation}
which will be relevant below.
Comparing with the $\ev{\vert\chi\vert^2}$ contribution in \cref{scalareqback},
\begin{equation}
\frac{\frac{H\dot{\phi}_0}{f^2}\ev{\vert\chi\vert^2}}{\frac{im^2}{f} \ev{\chi^2-\chi^{*2}}  } \simeq \frac{H\rho^3}{m^4}\simeq\frac{1}{8\xi}\lesssim1\, ,
\label{ratio2}
\end{equation}
which using \cref{xiapp} turns out to be smaller than unity.

Similarly to \eqref{modchi2-2}, \eqref{chipluschi} and \eqref{eqschiphibis-last2}, one can straightforwardly obtain a dimensional-analysis estimate  for  all the other operators in the equations \eqref{Friedmann1}--\eqref{scalareqback}.
In the end, the Friedmann equations reduce to
\eq{
3\Mpl^2H^2   \simeq \frac{\dot{\phi}_0^2}{2} +  V + \frac{\e^{2\pi\xi}}{4\pi^2}M^4,
\label{Friedmann1b}
}
\eq{
-\Mpl^2\dot H   \simeq \frac{\dot{\phi}^2_0}{2} + \frac{\e^{2\pi\xi}}{4\pi^2}M^4 .
\label{Friedmann2b}
}
For the inflaton background equation \eqref{scalareqback}, we find that the leading contribution comes from the last term in \eqref{scalareqback} (see eqs.~\eqref{ratiochi2chi2} and \eqref{ratio2}). As a result, it reduces to
\eq{
3H\dot\phi_0 + V'  + \frac{\e^{2\pi\xi}}{2\pi^2}\frac{m^4}{f} \simeq 0 \,.
\label{scalareqbackb}
}
Note that, from \cref{posO}, it follows that the sign of the last term in \eqref{scalareqbackb} is consistent with the sign of $3H\dot{\phi}_0$, as it should be in order for it to be interpreted as a standard friction in the inflation dynamics.\footnote{\label{foot:sign}{Recall that we have been working under the assumption that $\dot{\phi}_0>0$. However, from the symmetry $\{\phi\rightarrow-\phi,\chi\rightarrow\chi^*\}$ of the action \eqref{action}, it automatically follows that the same holds true even if we had chosen to work with the opposite convention $\dot{\phi}_0<0$. In fact, by definition, performing  the transformation $\{\phi\rightarrow-\phi,\chi\rightarrow\chi^*\}$ nothing can change. Therefore, since the left-hand side of \eqref{eqschiphibis-last2} flips sign, its right-hand side must flip sign as well, i.e.,~it must be odd in $\dot{\phi}_0$.}}

As anticipated, we focus on the large-backreaction regime $H\dot\phi_0 \lesssim \e^{2\pi\xi}m^4/(2\pi^2 f)$, in such a way that we can disregard $H\dot\phi_0$ in \eqref{scalareqbackb}. As a result, we obtain
\begin{equation}
2\pi \xi \simeq \ln \bigg\vert \frac{2\pi^2 f V'}{m^4}\bigg \vert \, .
\label{eq:log}
\end{equation}
On the other hand, we require the energy density to be dominated by the inflaton potential, like in the standard slow-roll paradigm, i.e.,
\begin{equation}
V \gg \dot{\phi}_0^2 \, , \, \,    \frac{\e^{2\pi\xi}}{4\pi^2}M^4 
\end{equation}
in such a way that $\Mpl^2H^2\simeq V$ and $\varepsilon\equiv -\dot H/H^2\ll 1$.

\subsection{Exponential Growth and Attractor Behavior}
\label{sec:attractor}
As a result of the instability, the amplitude of one of the two modes of the  $\chi$ field grows exponentially during the intermediate phase $t_1<t<t_2$. After $t_2$, the amplitude of the unstable mode gets enhanced by a factor $\e^{\pi\xi}$, with $\xi $ defined in eq.~\eqref{xi-app}. 
Using the hierarchy \eqref{hierarchy}, the exponent $\xi$ scales approximately as\footnote{See also \cref{app:xi}. Note that in \eqref{sec:xi} we reinstated the inflaton velocity $\dot{\phi}_0$.}
\begin{equation}
\xi \simeq  \frac{m^4 f}{8H \dot{\phi}_0 M^2}  \, .
%= \frac{m^4}{8H \rho M^2}\,.
\label{sec:xi}
\end{equation}
 A necessary condition for  the system to evolve toward an attractor is that $\xi $ should be a monotonically increasing function of the inflaton velocity, i.e.,~the friction and the energy injected into the $\chi$ sector should grow as $\dot{\phi}_0$ grows.
Using the expansions \eqref{M2m2expansion}, it is easy to choose the couplings $c_n$ and $c'_n$ in such a way that $\D \xi/\D \dot{\phi}_0 >0$ for values of the parameters that satisfy the condition \eqref{hierarchy}. In the following, we will  always work under this assumption of monotonicity of $\xi(\dot{\phi}_0)$.

%%%%%%%%%%%%%%%%%%%%%%%%%%%%%%%%%%%%%%%%%%%%%%%%%%%

\section{Perturbation Theory and Power Spectrum}
\label{sec:perturbations}

\subsection{Inflaton Perturbations}
\label{sec:ptfriction}

We next expand the fields in perturbations and study their dynamics around the inflationary background. 
We will neglect for  simplicity the mixing with gravity and  focus on the scalar perturbations only. In this section we will study linear perturbations and postpone the discussion of beyond-linear-order perturbations to \cref{sec:nongaussianity}.

Defining the inflaton fluctuation as $\varphi (t,\bfx)\equiv \phi(t,\bfx)- \phi_0(t)$ and using \cref{phi-eq}, the linearized equation for $\varphi$ can be schematically written as, 
\begin{equation}
\mathcal{D}  \varphi (t,\bfx) = \mathcal{O}[\chi; t]-\ev{\mathcal{O}}_{\vp=0}
%\delta \mathcal{O}[\chi; t] \, ,
\label{eqvarphi}
\end{equation}
where the differential operator $\mathcal{D}$ is given by
\begin{equation}
	\begin{split}
		\mathcal{D}  \equiv & -\left[  1+ \frac{\langle\vert \chi\vert^2\rangle}{\rho^2 f^2} \left( M_X^2 -2 \rho^2\right)-  \frac{1}{2\rho^2 f^2}\langle \chi^2+\chi^{*2}\rangle m_X^2 \right]\Box+V''
		\\
		& +\left(\frac{\langle\vert \chi\vert^2\rangle}{\rho^2 f^2} M_{XX}^2-  \frac{1}{2\rho^2 f^2}\langle \chi^2+\chi^{*2}\rangle  m_{XX}^2  \right)\frac{1}{a^3}\p_t(a^3\p_t)
		- \frac{i}{\rho f^2}\langle \chi^2-\chi^{*2}  \rangle m_X^2\partial_t\,,   
	\end{split}
	\label{Dop}
\end{equation}
where we have kept only the leading contributions in the quasi-de Sitter approximation.
For instance, we neglected the slow-roll suppressed terms $\ddot{\phi}_0\ll H \dot{\phi}_0$, $\langle \partial_t \vert \chi\vert^2\rangle \ll H \langle\vert \chi\vert^2\rangle$, and $\langle \partial_t(\chi^2+\chi^{*2})\rangle \ll H \langle \chi^2+\chi^{*2}\rangle$. Following the discussion in \cref{sec:computationexpvalues} and  using the hierarchy \eqref{hierarchy}, the operator $\cal D$ can be significantly simplified:\footnote{This will become more clear in section~\ref{localnn} below. To see this explicitly, one can consider the limit $2\pi\xi\gg 1$. From eqs.~\eqref{modchi2-2}, \eqref{ratiochi2chi2}, and \eqref{eqschiphibis-last2} it follows in fact that 
	\begin{equation}
		\frac{\langle\vert \chi\vert^2\rangle}{ f^2} \simeq \frac{1}{32\pi} \frac{\gamma/H}{\xi^2} \ll1 \, ,
		\quad
		\frac{m_X^2}{\rho^2 f^2}\langle \chi^2+\chi^{*2}\rangle \ll \frac{\langle\vert \chi\vert^2\rangle}{ f^2} \ll1  \, ,
		\quad
		\frac{i}{\rho f^2H}\langle \chi^2-\chi^{*2}  \rangle m_X^2 \simeq \frac{\gamma/H}{2\pi\xi} \ll1\, ,
\end{equation} 
where $\gamma$ is defined in \cref{gamma}.
However, analogous conclusions hold also in the other cases mentioned in section~\ref{localnn} that allow to achieve a local dynamics in the system.} 
\begin{equation}
	\mathcal{D}\varphi \simeq \ddot{\varphi}
	+\left(3H-\frac{i}{\rho f^2}\langle \chi^2-\chi^{*2}  \rangle m_X^2\right) \dot{\varphi}
	-\left(  \frac{\partial_i^2}{a^2}  -  V'' \right) \varphi   \, ,
	\label{Dsimple}
\end{equation}
with the value of $\ev{\chi^2-\chi^{*2}}$ given in \cref{eqschiphibis-last2}. To get  \eqref{Dsimple} we have assumed that $M^2_{XX}\sim M^2_X\sim M^2$ and $m^2_{XX}\sim m^2_X\sim m^2$.

The quantity $\mathcal{O}[\chi; t]$ on the right-hand side of \cref{eqvarphi} is a composite operator that depends quadratically on $\chi$, and acts as a source term in \eqref{eqvarphi}. More concretely, it can be seen from \cref{phi-eq} that there are five different quadratic combinations of $\chi$ contributing to $\mathcal{O}[\chi; t]$:  $\vert\chi\vert^2$, $(\chi^2+\chi^{*2})$, $\partial_t\vert\chi\vert^2$, $\partial_t(\chi^2+\chi^{*2})$ and $(\chi^2-\chi^{*2})$. 
For each of them we can identify two types of contributions~\cite{LopezNacir:2011kk}:
\begin{equation}
\mathcal{O}-\ev{\Oc}_{\vp=0} = \delta\mathcal{O}_S + \delta\mathcal{O}_R \, .
\label{deltaOsplit}
\end{equation}
$\delta\mathcal{O}_S\equiv [\mathcal{O}-\ev{\mathcal{O}}]_{\varphi=0}$ is the stochastic, or noise, contribution and represents intrinsic inhomogeneities in $\mathcal{O}$ that would be present even in the absence of $\varphi$ perturbations, i.e., deviations from the expectation value due to quantum fluctuations of the ADOF. The two- and three-point correlation functions induced by $\delta\mathcal{O}_S$  will be studied in sections~\ref{sec:inhomogeneoussolution} and \ref{sec:intrinsicNG} respectively (see also \cref{app:2pt}).  
On the other hand, $\delta\mathcal{O}_R$ is the response contribution, which corresponds to the change induced by inflaton perturbations, originating from the fact that the fields $\phi$ and $\chi$ are coupled.
In other words, $\delta\mathcal{O}_R$ can be expressed as a functional of $ \varphi$.\footnote{Note that, since Lorentz invariance is broken only spontaneously by $\dot{\phi}_0$ (and there are no other sources of breaking) and since $\phi$ is a shift-symmetric scalar, $\mathcal{O}$ can be unambiguously covariantized. Therefore, to the leading order in derivatives, it is a functional of the Lorentz scalar $\partial_\mu\phi\partial^\mu\phi$ \cite{LopezNacir:2011kk}. Thus $\delta\mathcal{O}_R$ depends on $\varphi$  through the combination $\dot\phi_0\dot\varphi+\dot\varphi^2/2-(\partial_i\varphi)^2/(2a^2)$. We will come back to this in \cref{sec:nongaussianity}.\label{nonLorentz}}
More precisely, at  linear level and for long-mode inflaton perturbations, the response $\delta\mathcal{O}_R$ simply corresponds to the variation of the expectation value $\ev{\Oc}$ due to a change in the inflaton velocity.
It can be written as an integral over the history of $\dot\vp$ (see \cref{nonLorentz}), subject to some kernel $K_{\Oc}(t,t')$:
\eq{
\delta\Oc_R(t,\vec k)=\delta\ev{\Oc}(t,\vec k)=\int\dd{t'}K_{\Oc}(t,t')\,\dot\vp_{\vec k}(t')\,.
\label{dOR}
} 
The integral in \cref{dOR} depends  in principle on higher derivatives of $\vp$ as well, but those are suppressed as long as we restrict ourselves to slowly-varying  and long-wavelength inflaton perturbations, compared to the instability scale. In this limit, it is possible to extend the WKB solution for $\chi$ to include effects of $\vp$.\footnote{The extension of the WKB procedure is immediate for homogeneous inflaton perturbations: we just replace $\phi_0\to\phi_0+\vp(t)$. However, for inhomogeneous perturbations one has to resort to \cref{eq:chi1} and apply general eikonal methods~\cite{Weinberg:PhysRev.126.1899}.} Then, one can use it to compute $\delta\Oc_R$ by expanding the expectation value of $\Oc$ to linear order in $\vp$.  
Note that the particular form of the kernel  depends crucially on the structure of the operator in terms of $\chi$, i.e.,~the five different combinations listed above. 

Given the above discussion, \cref{eqvarphi} is thus in general an integro-differential equation for $\vp$, sourced by the stochastic fluctuation $\delta\Oc_S$.\footnote{Note that, even at linear order in the inflaton perturbation, this equation contains a convolution over spatial momenta, which is  hidden inside the kernel $K_{\Oc}$. This is a consequence of the fact that $\chi$ does not have a zero mode at the level of the background. Note that this convolution remains even for spatially homogeneous inflaton configurations.}
However, a simplification can be made so that we do not need to deal with \cref{eqvarphi} in its general form. As we will argue in the following subsection, it is possible to consider a situation in which only one particular operator, $(\chi^2-\chi^{*2})$, on the right-hand side of \cref{eqvarphi} is relevant. For this operator the response is local, i.e., the kernel $K_{\Oc}$ is approximately proportional to a delta function, and therefore \cref{eqvarphi} is an ordinary differential equation.          

\subsection{Local vs.~Nonlocal Response}
\label{localnn}

Among the five different operators that appear on the right-hand side of \cref{eqvarphi}, we will argue that $\vert\chi\vert^2$ and $(\chi^2+\chi^{*2})$ behave in a similar way, and likewise $\partial_t\vert\chi\vert^2$ and $\partial_t(\chi^2+\chi^{*2})$. Therefore, we shall focus just on the three operators $\vert\chi\vert^2$, $\partial_t\vert\chi\vert^2$, and $(\chi^2-\chi^{*2})$.
One main qualitative difference among them is that, as it follows from section~\ref{sec:computationexpvalues} and figure~\ref{vevO}, the expectation value of the first one is sensitive to modes with momenta in a broad range of order of the Hubble scale, while the latter two dominantly receive contributions only from modes localized on a parametrically shorter range of scales, around the instability window. 
This distinction is physically relevant, because that is what discriminates between local and nonlocal behaviors.

To understand this, let us take the expectation value of these operators on the background solution $\dot{\phi}_0(t)$ and imagine slightly perturbing the inflaton velocity by adding a homogeneous fluctuation $\dot\vp(t)$.
Now consider for instance $\langle\vert\chi\vert^2\rangle$, which can be computed using \cref{modchi2-1} with the mode functions including the effect of the slowly varying inflaton velocity.
Every mode in the integration of \cref{modchi2-1} is exponentially sensitive to the value of the inflaton velocity around the time the mode becomes unstable, i.e., the time of crossing the turning points. 
Therefore, one could interpret the integral over $k$ in \cref{modchi2-1} as an integral over the history of the inflaton velocity. 
However, as shown in figure~\ref{vevO}, the result is sensitive to a wide range of scales. This suggests that the response of $|\chi|^2$ cannot be described in a local approximation. In contrast, $\ev{-i(\chi^2-\chi^{*2})}$ is expected to be local: as evident from the right panel in figure~\ref{vevO}, the main contribution to the result comes from a narrow range of scales around the instability window, which is parametrically shorter than a Hubble time.      

What we just described can be seen explicitly in figure~\ref{responseO},\footnote{The reason why we get an undershoot in the response plots in figure~\ref{responseO} is that, in order to avoid expensive numerical calculations, the separation between the instability scale and the Hubble scale is taken not huge. Therefore, in our code,  a perturbation with a wavelength $\lambda$ fraction of a  Hubble time is not precisely a long mode. We expect that in the limit $H/m\to0$ responses become more and more similar to a step function, like the driving perturbation.} where we compare the responses of $|\chi^2|$ and $(\chi^2-\chi^{*2})$ to a change in the inflaton velocity $\dot{\phi}_0$. One could easily imagine that this result is extendable to the case of long-wavelength, but {\it inhomogeneous}, perturbations, although this would require a more sophisticated analysis.

The situation with the operator $(\chi^2+\chi^{*2})$ is similar to the one of $|\chi|^2$. 
A similar plot as in figure~\ref{vevO} for the $(\chi^2+\chi^{*2})$ operator reveals that there are oscillations in $k/(aH)$ around a nonzero average  after the instability band. This means that $(\chi^2+\chi^{*2})$  has a nonlocal response as well. On the other hand,  for both $\p_t|\chi|^2$ and $\p_t(\chi^2+\chi^{*2})$, the oscillations happen around zero---very similarly to the case of $(\chi^2-\chi^{*2})$ shown in figure~\ref{vevO}---thanks to the time derivative which eliminates the nonzero average. Therefore, both $\p_t|\chi|^2$ and $\p_t(\chi^2+\chi^{*2})$ have local responses.  
 
\fg{
\includegraphics[width=0.49\textwidth,trim={0cm 0.5cm 1cm 1.5cm},clip]{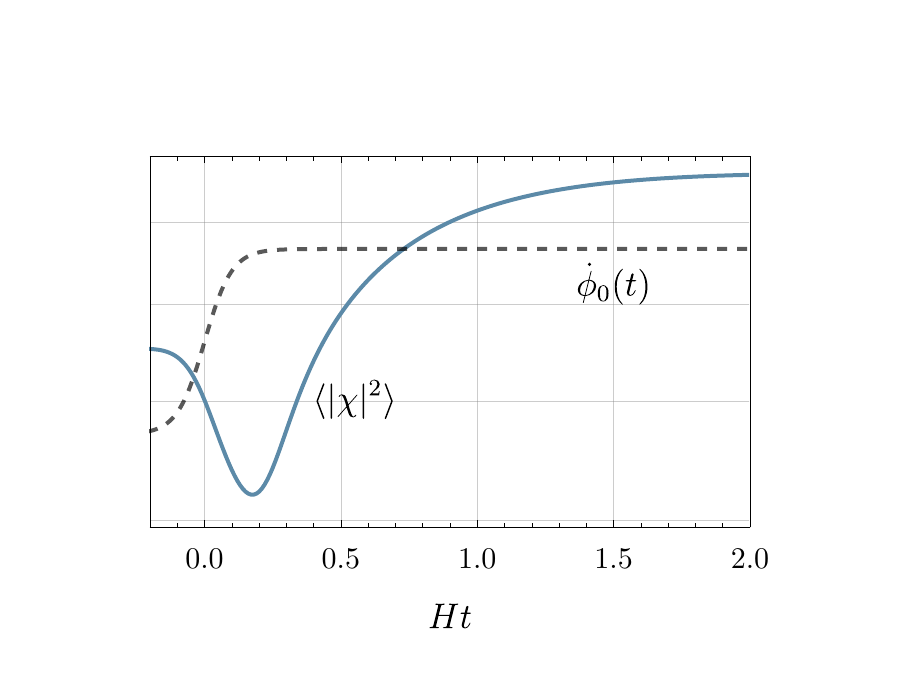}
\includegraphics[width=0.49\textwidth,trim={0cm 0.5cm 1cm 1.5cm},clip]{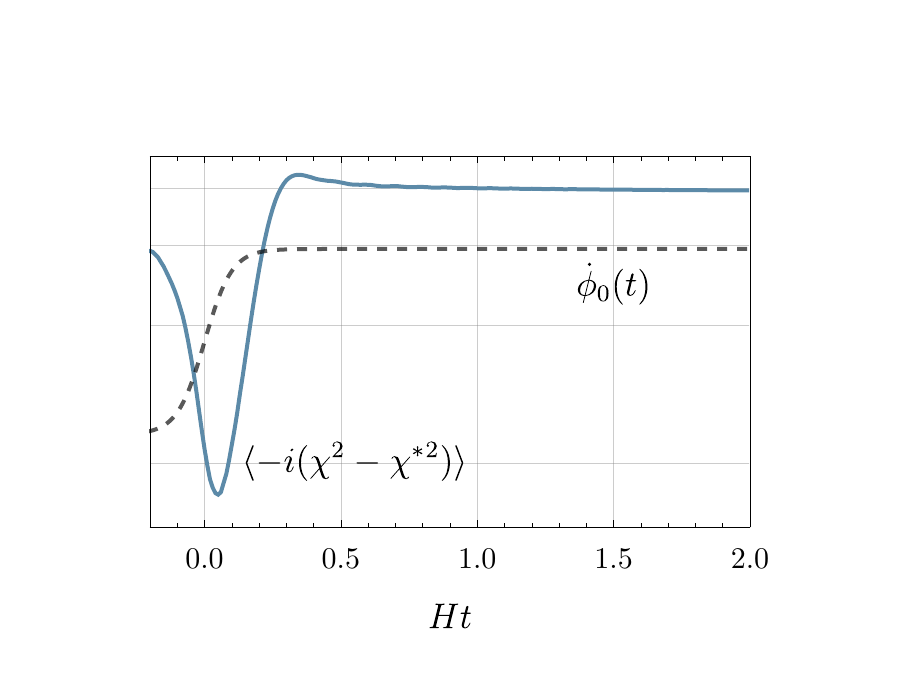}
\caption{Numerical solutions for the expectation values $\ev{|\chi|^2}$ (left) and $\ev{-i(\chi^2-\chi^*{}^2)}$ (right), assuming that the inflaton velocity (dashed line) has a step $\dot\vp\propto\tanh(t/\lambda)$ with $\lambda=0.1H^{-1}$. While $\ev{|\chi|^2}$ requires more than a Hubble time to reach the asymptotic value, $\ev{-i(\chi^2-\chi^*{}^2)}$ responds to the step function in a fraction of a Hubble time. This confirms locality of the response for the latter as opposed to the former operator. In the plots, units of the vertical axes are arbitrary.}
\label{responseO}
}

Having a nonlocal response in the model has some negative aspects. First of all, it makes an analytic treatment very complicated. More importantly, it is very likely that the system would exhibit a resonant instability for the inflaton perturbations. The situation  would then be similar to the case of axion inflation where such instabilities are present, as recently illustrated in~\cite{Domcke:2020zez,Caravano:2022epk,Peloso:2022ovc,Figueroa:2023oxc}.

Keeping the above discussion in mind, we now describe some possible ways in which we can suppress nonlocal response in the system. The first option is to take $2\pi\xi\gg 1$, while keeping $\ev{\chi^2-\chi^*{}^2} m^2/(f H \dot{\phi}_0) \simeq \e^{2\pi\xi}m^4/(H \rho f^2)$ fixed.
This suppresses parametrically the nonlocal response of $\frac{H\dot{\phi}_0}{f^2}\vert\chi\vert^2$ compared to $\frac{im^2}{f}(\chi^2-\chi^{*2})$ in the linearized \cref{eqvarphi}. 
As a byproduct, in this limit the response of the (local) operator $\partial_t\vert\chi\vert^2$ is also negligible.\footnote{Recall that $\langle \partial_t\vert\chi\vert^2\rangle$ is slow-roll suppressed. This implies in particular that, in linear response theory, the operator $\frac{\rho}{f} \partial_t\vert\chi\vert^2$   will not induce any $\dot{\varphi}$ term in the equation \eqref{eqvarphi}. It may however, and in general will, generate a $\ddot{\varphi}$ term, which would  correspond to a modification to the sound speed $c_s$. Nevertheless, using the WKB solution \eqref{fullWKB-main} and the hierarchy \eqref{hierarchy}, one finds that $\delta c_s^2 \ll 1$ in the limit $2\pi\xi\gg1$.} 
As a result,  $(\chi^2-\chi^{*2})$  becomes the leading source, not only for the background in~\cref{scalareqbackb}, but also for perturbations in~\cref{eqvarphi}.

Alternatively,  one could allow for a moderate `fine tuning'. Choosing  the couplings in \eqref{M2m2expansion} in such a way that $M_X^2=2\rho^2$ allows in principle to remove the nonlocal operator $|\chi|^2$ (as well as $\partial_t\vert\chi\vert^2$) from \cref{phi-eq} altogether.\footnote{Note that the other nonlocal operator $(\chi^2+\chi^{*2})$ is already parametrically smaller than $(\chi^2-\chi^{*2})$. Indeed, comparing the two terms in \eqref{scalareqback} yields
\begin{equation}
\frac{\frac{m^2_X}{\rho^2f^2} \ev{\chi^2+\chi^*{}^2}H\dot{\phi}_0 }{\frac{m^2}{f} \ev{\chi^2-\chi^*{}^2}} \sim \frac{H}{2\rho} \ll 1  
%\lesssim \frac{1}{8\xi} \left( \frac{m}{\rho} \right)^2 \, ,
\end{equation}
and so is its response contribution to the right-hand side of \eqref{eqvarphi}.
Note that the tuning $M_X^2=2\rho^2$ is however not fully robust against quantum loops, because the tree-level couplings $c_n$ and $c_n'$ in \eqref{M2m2expansion} get order-one corrections from loop diagrams in the theory. 
}

Another possibility is to couple the ADOF $\chi$ to some other additional fields,  which $\chi$ can decay into on time scales much shorter than $H^{-1}$, removing all nonlocality~\cite{LopezNacir:2011kk}. 
In this case, one has to additionally require that such a coupling does not alter significantly the process of $\chi$ production triggered by the inflaton. 
This can be done by properly choosing the scale associated with this coupling that controls the decay rate. 

For all these scenarios, the result is that the response of the system is local in time and  we can safely neglect all but the $(\chi^2-\chi^{*2})$ operator to compute $\delta\mathcal{O}_R$. A similar conclusion holds for $\delta\mathcal{O}_S$.\footnote{One can check a posteriori by repeating the analysis of \cref{sec:inhomogeneoussolution} that, in the local approximation and under the assumption of the hierarchy \eqref{hierarchy}, the $\varphi$ correlation functions induced by $\vert\chi\vert^2$, $(\chi^2+\chi^{*2})$, $\partial_t\vert\chi\vert^2$ and $\partial_t(\chi^2+\chi^{*2})$ are suppressed compared to those sourced by $(\chi^2-\chi^{*2})$.}

The local approximation will be appropriate for the study of cosmological perturbations, but it will break down if one is interested in inflaton perturbations that are much shorter (and faster) than Hubble.  For instance, if one starts from inhomogeneous initial conditions, short-scale perturbations may be present and not amenable of a local description. In particular, these short perturbations may trigger oscillations similar to the ones discussed in the case of gauge-field instability. However, these oscillations will be eventually suppressed when modes become sufficiently long that the local description is appropriate, as we will discuss momentarily. The situation is qualitatively similar to what happens in non-dissipative models with an EFT cutoff: inflaton perturbations that are shorter than the cutoff cannot be described within the EFT, but this does not preclude the calculation of the cosmological correlators.

\subsection{Linear Equation and Friction Coefficient}

From the considerations of the previous section, it is possible to consider cases in which we can focus only on the single composite operator $(\chi^2-\chi^{*2})$ on the right-hand side of \cref{eqvarphi} for which treating the response as local is a good approximation. Thus, with a slight abuse of notation, we will just define
\begin{equation}
\mathcal{O} \equiv -i(\chi^2-\chi^{*2})\,. 
\label{opO}
\end{equation}
Using  eqs.~\eqref{phi-eq} and \eqref{Dsimple}, the linearized equation of motion for $\varphi$ reduces to
 \begin{equation}
\ddot{\varphi} + \left( 3H + \frac{m_X^2}{\rho f^2}\ev{\mathcal{O}} \right) \dot{\varphi} -\left(\frac{\partial_i^2}{a^2} - V'' \right) \varphi = 
-\frac{m^2}{f}(\delta\mathcal{O}_{S} + \delta\mathcal{O}_{R} ) \, ,
\label{varphi-eq}
\end{equation}
where the operator $\mathcal{O}$ is now given in \cref{opO}.
By virtue of the local approximation, its response $\delta\mathcal{O}_{R}$ can be easily computed as
\begin{equation}
\delta\mathcal{O}_{R} \simeq \frac{\partial \ev{\mathcal{O}}}{\partial\dot{\phi}_0}\dot\varphi + O(\varphi^2) \, .
\label{Ominusresponse}
\end{equation}
Hence, the $\varphi$ equation \eqref{varphi-eq} can  be rewritten in Fourier space as
\begin{equation}
\ddot{\varphi}_\bfk + (3H + \gamma )\dot{\varphi}_\bfk +\left(\frac{k^2}{a^2} + V'' \right) \varphi_\bfk = 
-\frac{m^2}{f}  \delta\mathcal{O}_{S}(\bfk)  \, ,
\label{varphi-eq-2}
\end{equation}
where the additional friction coefficient $\gamma$ is given by 
\begin{equation}
\gamma \simeq \frac{m_X^2}{\rho f^2}\ev{\mathcal{O}}  + \frac{m^2}{f}\frac{\partial \ev{\mathcal{O}}}{\partial\dot{\phi}_0} 
\simeq  \frac{\xi m^4 }{\pi M f^2}  \e^{2\pi\xi} \, ,
\label{gamma}
\end{equation}
and where we used \eqref{eqschiphibis-last2} and  $\partial \xi/\partial \dot{\phi}_0 \sim \xi/(M f)$. For later convenience, we can write \eqref{varphi-eq-2} in conformal time as
\begin{align}
\label{varphi-eq-2-conf}
	\varphi''_\bfk-\left(\frac{2}{\tau}+\frac{\gamma}{H\tau}\right)\varphi'_\bfk+\left(k^2+\frac{V''}{H^2\tau^2}\right)\varphi_\bfk = -\frac{m^2}{fH^2\tau^2}\delta\mathcal{O}_S(\bfk),
\end{align}
where primes denote derivatives with respect to the conformal time $\tau$.
In the following, we will focus primarily on the regime where the effect of dissipation due to  the coupling to the ADOF is larger than the standard Hubble friction, i.e., $\gamma \gtrsim 3H$. In the opposite regime $\gamma \ll 6 \pi \xi H$, see eq.~\eqref{scalareqbackb}, dissipation becomes also irrelevant for the background evolution: in this weak backreaction regime, the $\chi$ production is exponentially sensitive to the parameters and as such it tends to be very scale dependent.

Before proceeding let us comment on an implicit simplification we made so far.  We are neglecting metric perturbations and working in the so-called decoupling limit (see \cite{LopezNacir:2011kk} for a discussion of the decoupling limit in the presence of dissipation). This is possible in the limit in which the first Friedmann equation \eqref{Friedmann1b} is dominated by the potential, with all deviations (kinetic energy of the inflaton and stress-energy tensor of the ADOF) suppressed by $\varepsilon \ll 1$. Neglecting $\varepsilon$ corrections is not only a welcome simplification, but also crucial if one wants to stick with an approximation in which the effect of the ADOF can be taken to be local in time. Indeed the stress-energy tensor of $\chi$ redshifts with a timescale of order Hubble, so that the solution for the constraint variables $N$ and $N^i$ (in the ADM formalism) is sensitive to the past dynamics of $\chi$.  

We are now ready to solve the (inhomogeneous) equation \eqref{varphi-eq-2-conf}.
The full solution for  $\varphi$ is in general given by the superposition of a piece that solves the homogeneous equation and a particular solution. 
The former corresponds to the usual inflaton vacuum fluctuations, which are discussed in section~\ref{sec:vacuumfluctuations} below.
The particular solution can instead be thought of as a perturbation induced by two $\chi$ modes and is derived explicitly in section~\ref{sec:inhomogeneoussolution}.

\subsection{Vacuum Fluctuations}
\label{sec:vacuumfluctuations}
Here we study the homogeneous part of the solution to the \cref{varphi-eq-2-conf}. 
Note that this is not the same as the conventional equation for the perturbations with $H$ simply replaced by $\gamma$, since the modes still get redshifted by Hubble. 
In fact, ignoring the potential term and in the limit $\gamma\gg H$, one can see that freeze-out of modes happens at $k/a\sim\sqrt{\gamma H}$ rather than the naive expectation $k/a\sim\gamma$~\cite{LopezNacir:2011kk}.\footnote{Note that there are two scales in the problem described by the homogeneous equation \eqref{varphi-eq-2}. We have $k/a\sim \gamma$ which is the scale at which the friction of the damped oscillator \eqref{varphi-eq-2} starts becoming important. After this, $\gamma\dot{\varphi}$ becomes dominant over the kinetic energy $\ddot{\varphi}$: $\gamma \dot{\varphi}_\bfk + \frac{k^2}{a^2} \varphi_\bfk=0$. Freezing happens when the decay rate $\frac{k^2}{a^2} \frac{1}{\gamma}$ becomes of order $H$, i.e., $k/a\sim\sqrt{\gamma H}\ll \gamma$. 
} 

We will see below that, as $\gamma$ becomes larger, the homogeneous solution becomes less important  than the particular solution \cite{LopezNacir:2011kk}.
We first note that the homogeneous solution to~\cref{varphi-eq-2-conf} can be generally written as
\eq{
\vp_k=\tau^\beta\left[A \,  {\sf H}^{(1)}_\alpha(-k\tau)+B \, {\sf H}^{(2)}_\alpha(-k\tau)\right]\,,
}
for constant $A$ and $B$, and where we defined 
\begin{equation}
	\alpha \equiv \sqrt{\left(\frac{3}{2}+\frac{\gamma}{2H}\right)^2-  \frac{V''}{H^2} } \,,\quad\quad \beta\equiv\frac{3}{2}+\frac{\gamma}{2H}\,.
\label{alphabeta}
\end{equation}
In this section the effect of the potential will be immaterial, so we shall replace $\beta\simeq\alpha$  in the following. 
Since the system is dissipative one must match the initial conditions to the Bunch--Davies vacuum at some early time $k|\tau_0|\gg\gamma/H$. 
Demanding the correct behavior around $\tau_0$ implies that $A=0$ and $B\sim \tau_0^{-\gamma/2H}$. Therefore, the asymptotic value of the solution in the limit $\tau\to0$ is
\eq{
\vp_k(\tau\to0)\propto\left(\frac{k\tau_0}{\gamma/H}\right)^{-\frac{\gamma}{2H}}\e^{-\frac{\gamma}{2H}}\,,
}
where we have used the fact that $\gamma\gg H$. 
As a result, increasing the friction will exponentially damp the homogeneous solution. 
We will ignore the vacuum fluctuations in the rest of the analysis; note that vacuum fluctuations may be subdominant even for small friction $\gamma \ll H$, i.e.~depending on the parameters it is possible to have negligible friction while the dominant source of fluctuations comes from the noise.
  
\subsection{Stochastic Noise and the Induced Power Spectrum}
\label{sec:inhomogeneoussolution}

The particular  solution to the inhomogeneous equation \eqref{varphi-eq-2-conf} can be derived using standard Green's function methods. 
We can write the solution in momentum space as
\begin{equation}
\varphi(\tau,\bfk)= -\frac{m^2}{f}\int \D\tau' G_k(\tau,\tau') a(\tau')^2 \delta\mathcal{O}_{S}(\tau',\bfk)  \, ,
\label{varphiinh}
\end{equation}
in which $G_k(\tau,\tau')$ is the retarded Green's function of \cref{varphi-eq-2-conf}. To leading order in the quasi-de Sitter approximation, it can be written in terms of the Bessel functions of first and second kind,\footnote{The retarded Green's function solution \eqref{greenssolution2} can be also equivalently written in terms of Hankel functions as
\begin{equation}
G_k(\tau,\tau') = 
\frac{\pi}{4i} \tau \left(\frac{\tau}{\tau'}\right)^{\beta-1} \left[ {\sf H}^{(1)}_\alpha (-k\tau){\sf H}^{(2)}_\alpha (-k\tau') - {\sf H}^{(2)}_\alpha (-k\tau){\sf H}^{(1)}_\alpha (-k\tau') \right]\Theta(\tau-\tau') \, .
\label{greenssolution3}
\end{equation}
}
\begin{equation}
G_k(\tau,\tau') = 
\frac{\pi}{2} \tau \left(\frac{\tau}{\tau'}\right)^{\beta-1} \left[ {\sf Y}_\alpha (-k\tau){\sf J}_\alpha (-k\tau') - {\sf J}_\alpha (-k\tau){\sf Y}_\alpha (-k\tau') \right]\Theta(\tau-\tau') \, .
\label{greenssolution2}
\end{equation}
Here $\alpha$ and $\beta$ are defined in \cref{alphabeta} and $\Theta$ is the standard Heaviside theta-function, defined such that $\Theta(x)=0$ for $x<0$ and $\Theta(x)=1$ for $x\geq 0$. As mentioned above, we will assume $\vert V''\vert^{1/2} \ll\gamma+3H$ which implies $\alpha \simeq \beta$ to leading order. 

Given the inhomogeneous solution \eqref{varphiinh}, the inflaton two-point function sourced by $\mathcal{O}$ can be obtained  by simply computing the equal-time expectation value
\begin{multline}
\langle\varphi_\bfk (\tau)\varphi_{\bfk'} (\tau)\rangle
	= \frac{m^4}{f^2}\int \D\tau'\D\tau'' a(\tau')^2 a(\tau'')^2 \,  G_k(\tau,\tau')G_{k'}(\tau,\tau'')
\langle \delta\mathcal{O}_{S} (\tau',\bfk) 
   \delta\mathcal{O}_{S}(\tau'',\bfk')\rangle \, .
%\int_\bfx \e^{-i\bfk\cdot\bfx}   \int_\bfy \e^{-i\bfk'\cdot\bfy} \langle \delta\mathcal{O}_{S} (\tau',\bfx)  \delta\mathcal{O}_{S}(\tau'',\bfy)\rangle \, ,
\label{varphivarphi}
\end{multline}
Here $\langle \delta\mathcal{O}_{S} (\tau',\bfk)\delta\mathcal{O}_{S}(\tau'',\bfk')\rangle$ is the (unequal-time) stochastic noise of $\mathcal{O}$ in momentum space, which using  eq.~\eqref{quantsigma} is found to  be
\begin{equation}
 \langle \delta\mathcal{O}_{S} (\tau',\bfk) 
   \delta\mathcal{O}_{S}(\tau'',\bfk')\rangle
   =
  4{(2\pi)^3}\delta(\bfk+\bfk') \, a(\tau')^{-3} a(\tau'')^{-3} \int \frac{\D^3\bfp}{(2\pi)^3} \mathbb{F}_{\vert\bfp\vert,\vert\bfk-\bfp\vert}(\tau',\tau'')
   \, ,
\label{noisek}
\end{equation}
where we defined 
\begin{equation}
\spl{
\mathbb{F}_{p_1,p_2}(\tau',\tau'') \equiv \left[F_{p_1}(\tau')\cdot F_{p_1}^{\dagger}(\tau'')\right]_{11}&\left[F_{p_2}(\tau')\cdot F_{p_2}^{\dagger}(\tau'')\right]_{22}\\
&+ \left[F_{p_1}(\tau') \cdot F_{p_1}^{\dagger}(\tau'')\right]_{12}\left[F_{p_2}(\tau')\cdot F_{p_2}^{\dagger}(\tau'')\right]_{21} .
}
\label{defFF}
\end{equation}
Note that the noise \eqref{noisek} is correctly proportional to the momentum-conserving delta-function $\delta(\bfk+\bfk')$, as it should be from invariance under spatial translations.

Now, it is not hard to see that under the assumptions of~\cref{sec:localapprox} of a narrow and sub-Hubble instability band, the two-point function \eqref{noisek} is local in both time and space, i.e., it can be approximately written as proportional to a delta function $\delta(\tau'-\tau'')$ (locality in time) and it is independent of the momentum $k$ (locality in space). The dilation isometry of de Sitter implies that the two-point function is of the form 
\begin{equation}
 \langle \delta\mathcal{O}_{S} (\tau',\bfk) 
   \delta\mathcal{O}_{S}(\tau'',\bfk')\rangle
   \simeq {(2\pi)^3}
  \delta(\bfk+\bfk')  \,   \delta(\tau'-\tau'') \, H^4 \tau^4 \nu_{\mathcal{O}} \, ,
\label{noisek-2}
\end{equation}
where the overall coefficient $\nu_{\mathcal{O}}$ does not depend on time (up to slow-roll corrections) and momentum~\cite{LopezNacir:2011kk}.\footnote{If one writes this expression in the Minkowski limit using physical momentum $\bfp$ and cosmological time $t$, one gets
\begin{equation}
 \langle \delta\mathcal{O}_{S} (t',\bfp) 
   \delta\mathcal{O}_{S}(t'',\bfp')\rangle
   \simeq {(2\pi)^3}
  \delta(\bfp+\bfp ')  \,   \delta(t'-t'') \,  \nu_{\mathcal{O}} \, .
\label{noiseMink}
\end{equation}
Therefore $\nu_{\mathcal{O}}$ is the Minkowski normalization. We expect, and we will verify shortly, that this does not depend on $H$ but only on the mass scales $M$ and $m$. Of course $H$ appears in the exponential growth of the modes, since it is the scale that cuts off the instability.}
The reasoning that justifies locality in space and time goes schematically as follows. 
Plugging the explicit solution \eqref{fullWKB-main} in \eqref{noisek}, one can check that the integrand, for fixed say $\tau''$, is everywhere an oscillating function of $\tau'$ with a zero mean value, except when $\tau' $ is sufficiently close to $\tau''$. 
For $\tau'\sim\tau''$ there is a nontrivial overlap between the instability regions of the mode functions $F(\tau')$  and $F^\dagger(\tau'')$ in each term $[F(\tau')\cdot F^{\dagger}(\tau'')]$ in \eqref{defFF}, resulting in a net positive, non-oscillating exponential enhancement. This is the interval where the integral of $\mathbb{F}_{\vert\bfp\vert,\vert\bfk-\bfp\vert}(\tau',\tau'')$ gets its leading contribution. 
In addition, since the exponential amplification takes place on very short scales while we are interested in the late-time two-point function of $\varphi$ i.e., in the limit $k\tau\ll1$, we can approximate $\vert\bfk-\bfp\vert\simeq \vert\bfp\vert$. 
As a result,
\begin{equation}
\nu_{\mathcal{O}} \simeq   \frac{1}{a(\tau')^2}\int \frac{\D^3\bfp}{(2\pi)^3}  \frac{\mathbb{F}_{\vert\bfp\vert,\vert\bfp\vert}(\tau',\tau')}{\omega(p) \, a(\tau')}
\sim  \frac{\e^{4\pi \xi}}{4\pi^2} \frac{M}{m} \, ,
\label{FFintk0}
\end{equation}
where in the second equality we used that $\omega^2\simeq \frac{m^2}{2\rho^2}(k_{\text{ph}*}^2-\kp^2)$ close to the turning point $k_{\text{ph}*}=\sqrt{M^2-m^2}$ where the frequency vanishes,  and simple dimensional analysis.\footnote{Note that in estimating the integral \eqref{FFintk0} we are disregarding UV divergences. Since the leading contribution to the integral comes from a narrow region and is largely independent of what happens on much shorter scales, we can safely neglect the details of any regularization procedure needed to take care of the divergences.}
We provide a more detailed derivation of \eqref{FFintk0} in \cref{app:2pt}. 

By using \cref{noisek} in \eqref{varphivarphi} we get 
\begin{equation}
	\langle\varphi_\bfk (\tau)\varphi_{\bfk'} (\tau)\rangle
	\simeq  {(2\pi)^3} \delta(\bfk+\bfk') \frac{m^4}{f^2}\nu_{\mathcal{O}} 
	\int \D \tau' G_k(\tau,\tau')^2 \, .
	\label{phiphi2}
\end{equation}
Now, we are interested in the two-point function at late times $\tau\to0$. In this limit, the expression for the Green's function given in \cref{greenssolution2} simplifies as
\eq{
	G_k(\tau,\tau')\simeq \frac{1}{k}(-k\tau)^{\beta-\alpha}\Gamma(\alpha)2^{\alpha-1}\frac{{\sf J}_\alpha(-k\tau')}{(-k\tau')^{\beta-1}}\,,\qquad\tau\to0\,,
	\label{Gapp}
}
where the theta function in \cref{greenssolution2} is automatically satisfied if $\tau\to0$. To estimate the integral in \cref{phiphi2}, we can take the limit that the induced friction is dominant over the Hubble friction, $\gamma\gtrsim H$, and  set, to leading order in the slow-roll approximation, $\alpha \simeq \beta$.  Therefore by using \cref{Gapp} with $\beta\simeq\alpha$ the integral is~\cite{LopezNacir:2011kk}
\begin{equation}
	\int \D \tau' G_k(\tau,\tau')^2
	= \frac{1}{k^3}\frac{4^{\alpha-2} \Gamma (\alpha -1) \Gamma (\alpha )^2}{ \Gamma(\alpha-\frac{1}{2})\Gamma(2\alpha-\frac{1}{2})}\simeq \frac{1}{4k^3}\sqrt{\pi/(2\alpha-3/2)}\, ,
	\label{calcGreens}
\end{equation}
where the last expression is an approximation obtained from asymptotic expansion for large $\alpha$ which is also reliably good for $\gamma/H \gtrsim 1$. 
Putting everything together:
\begin{equation}
	\langle\varphi_\bfk (\tau)\varphi_{\bfk'} (\tau)\rangle
	\simeq {(2\pi)^3} \delta(\bfk+\bfk') \frac{1}{k^3}  \bigg(\frac{Mm^3}{16 f^2} \frac{\e^{4\pi\xi}}{\sqrt{\gamma/H}\pi^{3/2}}\bigg)
	 \, .
	\label{powerspectrumstoch1}
\end{equation}
As expected, the power spectrum \eqref{powerspectrumstoch1} scales as $k^{-3}$, consistently with the scale invariance of de Sitter.
We can thus estimate the induced power spectrum  for the curvature perturbation $\zeta = -H\varphi/\dot{\phi}_0$ as:
\begin{equation}
	\Delta_\zeta^2 \equiv \frac{k^3}{2\pi^2} \frac{H^2}{\dot{\phi}_0^2} \langle\varphi_\bfk \varphi_{-\bfk}\rangle'
	\simeq \frac{1}{32\xi^2}\left(\frac{\gamma}{\pi H}\right)^{3/2}\frac{H^4M}{m^5}\,, 
	\label{power}
\end{equation}
which is consistent with the observed value $\Delta_\zeta^2\simeq 2\cdot 10^{-9}$ for typical choices of the scales in our model.
Note that in \cref{power} the prime symbol $'$ on the two-point function means that we removed the momentum-conserving delta-function ${(2\pi)^3}\delta(\bfk+\bfk')$. 
In addition, we neglected deviations from scale invariance. As usual, the tilt of the spectrum \eqref{powerspectrumstoch1} can be computed by taking into account the time dependence of $\zeta$ at horizon crossing: it contains various contributions due to the time dependence of the parameters in eq.~\eqref{power}. Notice that, in the strong backreaction regime, the tilt is not enhanced by the exponential dependence in eq.~\eqref{powerspectrumstoch1}: indeed the exponential can be written in terms of the derivative of the potential as in eq.~\eqref{eq:log} and it is, as such, slowly varying. Conversely, in the weak backreaction regime, the scale dependence will be enhanced by $4\pi\xi$.

The dependence of the power spectrum on $H$, $\Delta_\zeta^2 \propto H^4$, is easy to understand and it captures most of the physics. A rough estimate of the power spectrum can be obtained comparing the typical fluctuation $\delta\mathcal{O}_{S}$ with the average value $\langle\mathcal{O}\rangle$ on the unperturbed solution. One is interested in fluctuations on scales that are close to horizon  crossing, $k \sim \tau^{-1}$, and averaged on a window of order unity in conformal times. Using the power spectrum in eq.~\eqref{noisek-2}, one sees that this procedure gives something proportional to $H^4$. This can also be understood intuitively as follows. $\delta\mathcal{O}_{S}$ receives contributions from many independent short modes, characterized by the scales $m$ and $M$, which are both much larger than $H$ and that we can take as equal for this qualitative discussion. The power spectrum will go as $1/N$ where $N$ is the number of independent contributions. Since we are averaging, both in space and in time, with a typical scale $H$, the number of independent contributions goes as $(H/M)^4$. The smallness of the power spectrum is due to the presence of many independent sources that contribute incoherently as discussed in the introduction. In contrast, the case of axion inflation in the strong-backreaction regime predicts large power for the fluctuations simply because the instability continues up to large scales \cite{Anber:2009ua}.

%%%%%%%%%%%%%%%%%%%%%%%%%%%%%%%%%%%%%%%%%%%%%%%%%%%

\section{Non-Gaussianity}
\label{sec:nongaussianity}

In this section, we extend the previous discussion to next order in perturbation theory and estimate the size of non-Gaussianity in the model \eqref{action}.
Let us expand \cref{phi-eq} to quadratic order in perturbations.
We neglect the contributions from the operators $|\chi|^2$ and $(\chi^2+\chi^*{}^2)$ which, under the assumptions of \cref{localnn} for a local dynamics, are suppressed compared to those from $\mathcal{O} \equiv -i(\chi^2-\chi^{*2})$.
We then have
\begin{multline}
- \nabla_\mu \nabla^\mu\varphi  + V'' \varphi  +\frac{1}{2} V'''\varphi^2 
+ \frac{1}{\rho f^2}\left[ m_X^2 \dot\varphi +  \left( m_X^2 + m_{XX}^2 \right) \frac{\dot{\varphi}^2}{2\rho f}- m_X^2 \frac{(\partial_i\varphi)^2}{2\rho f a^2} \right] \ev{\mathcal{O}}
\\
+ \frac{1}{f} \left( m^2 +m_X^2 \frac{\dot \varphi}{\rho f}  \right) \left(\delta\mathcal{O}_R +\delta\mathcal{O}_S \right)
 = 0 \, ,
\label{phi-eq-quad}
\end{multline}
where $\delta\mathcal{O}_R$ is a functional of $\varphi$, which must be expanded up to second order as explained below. 
From \cref{phi-eq-quad} we can identify three different types of sources of non-Gaussianity. 
These are the usual sources of non-Gaussianity of standard inflationary models. 
They are slow-roll suppressed and independent of the coupling to the ADOF, and we will therefore not consider them here. 
In addition, there are the non-Gaussianities induced by the non-Gaussian statistics of the noise fluctuations $\delta \mathcal{O}_S$. 
We will briefly discuss them in~\cref{sec:intrinsicNG} below, where we show that their contribution is typically  subdominant. Finally, there are the non-Gaussianities sourced by the nonlinear coupling between $\varphi$ and the ADOF. These are of two types: they either result \textit{i)}  from nonlinear terms of the schematic form $\varphi \delta \mathcal{O}_S$ in the equations of motion (which stem from the $X$ dependence of $m^2$), or \textit{ii)}  from the nonlinear dependence of $\delta \mathcal{O}_R$ on $\varphi$. 

To obtain $\delta\Oc_R$ beyond linear order we note that, since $\phi_0(t)$ is the only source of breaking of Lorentz symmetries, nonlinearly realized Lorentz symmetry and shift symmetry ensure that $\delta\mathcal{O}_R$ depends on $\varphi$, to the leading order in derivatives, only through the Lorentz scalar $\partial_\mu \phi \partial^\mu\phi$. In other words, we can write
\begin{equation}
\spl{
\delta \mathcal{O}_R  &\equiv  \mathcal{O}\big[\sqrt{X}\big] - \mathcal{O}\rvert_{\varphi=0} 
= \mathcal{O}\big[\sqrt{X}\big] -  \left(\ev{\Oc}\rvert_{\varphi=0} + \delta \mathcal{O}_S\right) \ \\ 
&= \langle {\cal O}\rangle\big[\sqrt{X}\big] - \ev{\Oc}\rvert_{\varphi=0} + \delta {\cal O}\big[\sqrt{X}\big]- \delta \mathcal{O}_S \, ,
}
\label{Ominusresponse-200}
\end{equation}
where in the second line by $\delta {\cal O}$ we mean the fluctuation of the operator ${\cal O}$ around its expectation value $\langle {\cal O}\rangle$ (we stress that these quantities will in general depend on $\varphi$). Since we are neglecting in this section non-Gaussianity in the noise, we can treat $\delta \mathcal{O}$ as a Gaussian variable. In the local approximation \eqref{noisek-2}, $\delta \mathcal{O}$ must thus be proportional to $\delta \mathcal{O}_S$ and we can write, to linear order in $\varphi$, 
\begin{align}
	\delta {\cal O}\big[\sqrt{X}\big] \simeq \left[ 1+ \frac{1}{2\ev{\delta \Oc^2}}\frac{\partial \ev{\delta \Oc^2}}{\partial\dot \phi_0}  \bigg\vert_{\varphi=0}\dot{\varphi}\right]\delta {\cal O}_S
	= \left[ 1+ \frac{1}{2\nu_{\mathcal{O}}}\frac{\partial \nu_{\mathcal{O} }}{\partial\dot \phi_0}\dot{\varphi}\right]\delta {\cal O}_S \, ,
\end{align}
where the constant of proportionality in front of $\dot \varphi$ can be fixed by matching the variance of $\delta \mathcal{O}$ and $\delta \mathcal{O}_S$. Plugging into \cref{Ominusresponse-200}, we get
\begin{equation}
\delta \mathcal{O}_R  \simeq \frac{\partial \ev{\mathcal{O}}}{\partial\dot{\phi}_0} \left( \dot\varphi - \frac{(\partial_i\varphi)^2}{2\dot{\phi}_0 a^2} \right)  + \frac{1}{2} \frac{\partial^2 \ev{\mathcal{O}}}{\partial\dot{\phi}_0^2} \dot{\varphi}^2  + \frac{1}{2\nu_{\mathcal{O}}}\frac{\partial \nu_{\mathcal{O} }}{\partial\dot \phi_0} \dot{\varphi}\delta\mathcal{O}_S + \dots \, ,
\label{Ominusresponse-2}
\end{equation}
which extends \cref{Ominusresponse} to second order in perturbations. 

We stress again that, once the $\dot\vp$ term in the response is present, the nonlinearly realized Lorentz symmetries guarantee that also a $-(\partial_i\varphi)^2/(2 \dot{\phi}_0 a^2)$ term with fixed coefficient is present in $\delta \mathcal{O}_R $ \cite{LopezNacir:2011kk}. 
This is not necessarily true for $\dot\vp^2$, as it is ``contaminated" by other contributions.\footnote{The situation is very similar to the EFT of inflation where the coefficient of $\dot\pi(\p_i\pi)^2$ term in the action, unlike $\dot\pi^3$, is fixed by the quadratic action.} 
As we will see, the last term in \eqref{Ominusresponse-2} will provide the leading contribution to non-Gaussianity if $\gamma/H$ is not very large. 

Plugging \cref{Ominusresponse-2} back into \cref{phi-eq-quad}, we find
\begin{multline}
	\ddot{\varphi} + (3H +\gamma)\dot \varphi - \frac{\partial_i^2\varphi}{a^2}  + V'' \varphi  +\frac{1}{2} V'''\varphi^2 
	- \gamma \frac{(\partial_i\varphi)^2}{2\rho f a^2}
	\\
	+ \frac{1}{\rho f^2} \left[ \frac{1}{2\rho f} (m_X^2 + m_{XX}^2) \ev{\mathcal{O}} + m_X^2 \frac{\partial \ev{\mathcal{O}}}{\partial \dot{\phi}_0}  + \frac{\rho f}{2}m^2 \frac{\partial^2\ev{\mathcal{O}}}{\partial \dot{\phi}_0^2} \right] \dot{\varphi}^2
	\\
	= -\frac{1}{f} \left( m^2 + m_X^2 \frac{\dot{\varphi}}{\rho f} \right)\delta \mathcal{O}_S -\frac{m^2}{2f}\frac{\partial \ln\nu_{\cal O}}{\partial{\dot{\phi}_0}}\dot{\varphi}\delta{\cal O}_S\, ,
	\label{phi-eq-quad--2}
\end{multline}
where the friction coefficient $\gamma$ can be read off from \cref{gamma}. 
Assuming $m_{XX}^2\sim m^2_{X}\sim m^2$, and using eq.~\eqref{eqschiphibis-last2} for $\ev{\Oc}$,  $\pdv{\ev{\Oc}}{\dot\phi_0}\simeq2\pi\xi\ev{\Oc}/\rho f$ and $\partial(\ln \nu_{\cal O})/\partial{\dot{\phi}_0} \simeq (4\pi \xi)/(\rho f)$,~\cref{phi-eq-quad--2} takes approximately the form
\eq{
\vp''+(2H+\gamma)a\vp'-\partial_i^2\vp+a^2V''\vp  \simeq \frac{\gamma}{2\rho f}[(\partial_i\vp)^2-2\pi\xi\vp'^2]-a^2\frac{m^2}{f}\left(1+2\pi\xi\frac{\vp'}{a\rho f}\right)\delta\Oc_S\,,
\label{infeqn2}
}
where we switched to conformal time. We have neglected the self-interaction term $V'''$ in the final form as it is not relevant in this analysis. 
To obtain non-Gaussianity we must solve this equation including the nonlinear terms which will be the topic of the next subsection. 

Before moving into the details, we fix the notation: we parametrize the three-point function of the comoving curvature perturbation $\zeta=-\frac{H}{\dot\phi_0}\vp$ (in the spatially flat gauge) as
\eq{
	\langle\zeta_{\vec k_1}\zeta_{\vec k_2}\zeta_{\vec k_3}\rangle={(2\pi)^3}\delta(\vec k_1+\vec k_2+\vec k_3)B(k_1,k_2,k_3)\,.
	\label{zeta3}
}
Ignoring deviations from scale invariance, we define the amplitude in the equilateral configuration as
\eq{
	f_{\rm NL}^{\text{eq}}\equiv\frac{10}{9}\frac{k^6B(k,k,k)}{(2\pi)^4\Delta_\zeta^4}\,,
	\label{fnldef}
}
where $\Delta_\zeta^2$ is the power spectrum given in \cref{power}. The shape  can be parametrized in terms of 
\eq{
	S(x_2,x_3)\equiv (x_2x_3)^2\frac{B(k_1,x_2k_1,x_3k_1)}{B(k_1,k_1,k_1)}\,,
	\label{shapebis}
}
where $x_2=k_2/k_1$ and $x_3=k_3/k_1$, and to avoid redundant configurations we restrict to the region $\max(x_3,1-x_3)\leq x_2\leq1$.

\subsection{Nonlinear Evolution of Perturbations}
\label{sec:nonlinearNG}

The primary source of non-Gaussianity is due to nonlinear terms in \cref{infeqn2}. To study their effect we can solve \cref{infeqn2} perturbatively, following Ref.~\cite{LopezNacir:2011kk}. The leading-order solution was already given in \cref{varphiinh} where we have neglected the vacuum fluctuations as discussed in \cref{sec:vacuumfluctuations}. At the next-to-leading order (NLO), the solution to the quadratic equation \eqref{infeqn2} can be analogously found  in terms of the Green's function:
\begin{equation}
\begin{split}
		\vp_{\vec k}^{\rm NLO}(\tau)=-\frac{1}{\rho f}\int\dd{\tau'}&G_k(\tau,\tau')\int\frac{\dd[3]{\vec p}}{{(2\pi)^{3}}}\bigg[
		 \frac{\gamma}{2}\,\vec{p}\cdot(\vec k - \vec p)\,\vp_{\vec p}(\tau')\vp_{\vec k- \vec p}(\tau')\\ &+\pi\xi\gamma \, \p_{\tau'}\vp_{\vec p}(\tau')\p_{\tau'}\vp_{\vec k- \vec p}(\tau')
+2\pi\xi a(\tau')\frac{m^2}{f}\p_{\tau'}\vp_{\vec k-\vec p}(\tau')\delta\Oc_S(\tau',\vec{p})\bigg]\,,
\end{split}
		\label{phinnl}
\end{equation}
where $\vp$ in the integrand is the leading-order solution  in \cref{varphiinh} and $G_k(\tau,\tau')$ is the Green's function \eqref{greenssolution2}. The three-point function from \cref{phinnl} can be written as
\eq{
\langle\zeta_{\vec k_1}\zeta_{\vec k_2}\zeta_{\vec k_3}\rangle=-\left(\frac{H}{\rho f}\right)^3\left[\langle\vp^{\rm NLO}_{\vec k_1}\vp_{\vec k_2}\vp_{\vec k_3}\rangle+  \{\vec k_1\leftrightarrow\vec k_2\}+\{\vec k_1\leftrightarrow\vec k_3\}\right]\,,
\label{zzzNLNG}
}
where in this Section we assume that the statistics of $\delta\Oc_S$ is Gaussian.
The three different quadratic operators on the right-hand side of \eqref{phinnl} contribute to non-Gaussianity and we will now study them separately.

Let us start by  looking at the $(\p_i\vp)^2$ term in the equation of motion \eqref{infeqn2} or equivalently the first term in brackets in \cref{phinnl}.  After straightforward  algebraic manipulations, the three-point function at late times ($\tau\rightarrow 0$) is given by
\begin{multline}
	B_{\scriptscriptstyle{(\p_i\vp)^2}}(k_1,k_2,k_3)=\frac{\gamma}{H}\left(\frac{H}{\rho f}\frac{m^2}{f}\right)^4\nu_{\mathcal{O}}^2\int\dd\tau_1 \dd\tau_2\dd\tau_3  \, G_{k_1}(0,\tau_1)G_{k_2}(0,\tau_2)G_{k_3}(0,\tau_3)\times\\
	\times \left[\vec{k}_2\cdot \vec k_3 \, G_{k_2}(\tau_1,\tau_2)G_{k_3}(\tau_1,\tau_3)+{\rm 2\,\, perms.}\right]  ,
	\label{F1}
\end{multline}
where we have used Wick's theorem to write the four-point function of the noise in terms of the two-point function given in \cref{noisek-2}. The amplitude in the equilateral configuration, defined in \cref{fnldef}, is given by
\eq{
	f_{\rm NL}^{\text{eq}}=-\frac{5}{12}\frac{\gamma}{H}\left(\frac{4^{\alpha-2} \Gamma (\alpha -1) \Gamma (\alpha )^2}{ \Gamma(\alpha-\frac{1}{2})\Gamma(2\alpha-\frac{1}{2})}\right)^{-2}\int_0^{\infty}\dd{y}\tilde{G}(0,y)\left(\int_y^\infty\dd{y'}\tilde{G}(0,y')\tilde{G}(y,y')\right)^2\,.
	\label{fNL1}
}
Here we have defined $\tilde{G}(-k\tau,-k\tau')\equiv k\,G_k(\tau,\tau')$ which only depends on the conformal time through the scale invariant combination $k\tau$, and we have changed the integration variable to $y=-k\tau$. The minus sign in \cref{fNL1} appears since in the equilateral configuration $\vec{k}_2\cdot \vec k_3=-k^2/2$ is negative.  
Notice that \eqref{fNL1} only depends on the ratio $\gamma/H$, which appears  through $\alpha=3/2+\gamma/(2H)$ both in the prefactor and in the Green's functions. In the limit of large $\gamma/ H$ it is possible to  estimate numerically \cref{fNL1} as
\eq{
f_{\rm NL}^{\text{eq}}\simeq-\frac{\gamma}{4H}\,,
\label{fNLgammaH}
}
in agreement with the general analysis of~\cite{LopezNacir:2011kk}.
The numerical plot of $f_{\rm NL}^{\text{eq}}$ as a function of $\gamma/H$ is shown in figure~\ref{fNLplot}. 
We stress that the expression \eqref{fNLgammaH} for $f_{\rm NL}^{\text{eq}}$ in terms of the friction is enforced by nonlinear realization of Lorentz symmetries~\cite{LopezNacir:2011kk}. This is reminiscent of the relation between non-Gaussianity and reduced speed of sound \cite{Cheung:2007st}. Note that also the sign is the same in the two cases.

The shape of the bispectrum $S(x_2,x_3)$ is parametrized in \cref{shapebis}. The numerical computation of $S(x_2,x_3)$ for the $(\p_i\vp)^2$ term is shown in figure~\ref{shapeplot1}. The dominant peak is in the equilateral configuration. The squeezed limit $k_3\ll k_1$, $k_2$, or equivalently $(x_2,x_3)\simeq(1,0)$, is vanishing in the limit of scale invariance. 
This is consistent with the fact that the model has a single clock that controls the end of inflation, and the produced $\chi$ particles do not affect that. It thus satisfies the single-field consistency relations~\cite{Maldacena:2002vr,Creminelli:2004yq,LopezNacir:2012rm}.
One can see an enhancement in the collinear configuration, which does not grow  for $\gamma \gg H$, as one may have instead expected:  for $\gamma \gg H$ modes freeze well inside the horizon when the metric can be taken to be Minkowski; in this limit one has to conserve both energy and momentum and one gets correlation only among modes which are collinear. This logic however is not correct. Even if the modes freeze in Minkowski space, their dynamics before freezing is dominated by the friction term; they are not oscillating and energy is not conserved since it is lost in the friction due to $\chi$ production. 
\fg{
	\includegraphics[width=0.65\textwidth]{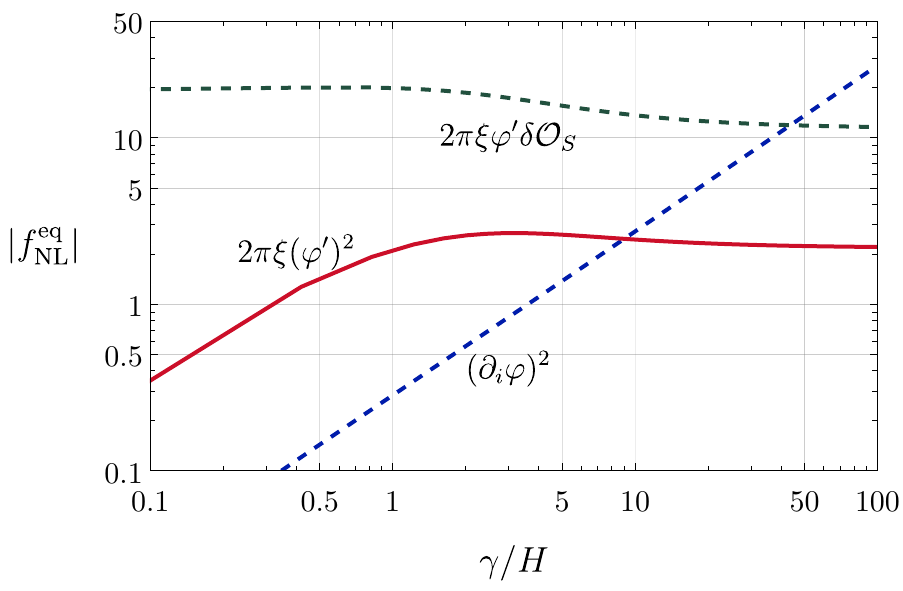}
	\caption{Absolute value of the amplitude \eqref{fnldef} for the three-point correlator in the equilateral configuration, computed from each quadratic operator in \cref{infeqn2}, as a function of the induced friction.   %as parameterized in \cref{fnldef} originated from nonlinear evolution as a function of the induced friction. 
	A solid (dashed) curve means that $f_{\rm NL}^{\rm eq}$ is positive (negative).
	%We plot the positive (negative) values with solid (dashed) curves.
	For the contribution from $\vp'^2$ and $\vp'\delta\Oc_S$ we need to specify the value of the growth exponent, for which we chose  $\xi=2$. 
}
	\label{fNLplot}
}
\fg{
	\includegraphics[width=0.7\textwidth,trim={0 2cm 0 3cm},clip]{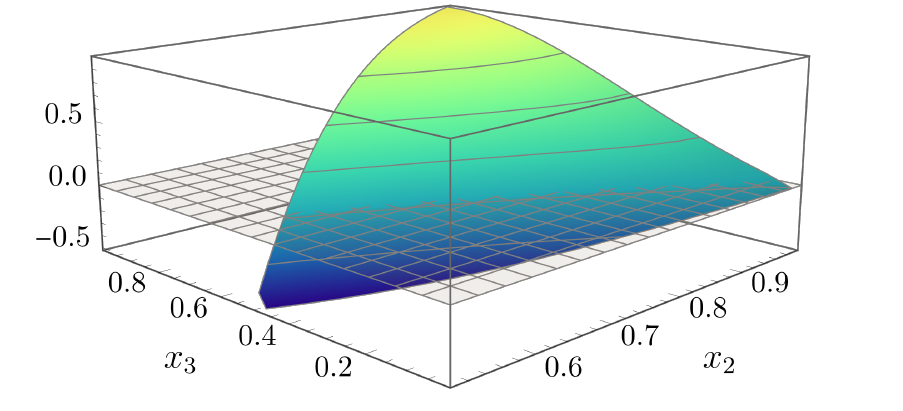}
	\caption{Shape function $S(x_2,x_3)$ defined in \cref{shapebis} for the nonlinear term $(\p_i\vp)^2$. The two parameters are $x_2=k_2/k_1$ and $x_3=k_3/k_1$. The plot is produced for the value $\gamma/H=5$.}
	\label{shapeplot1}
}

The next nonlinear term in \cref{infeqn2} is $\vp'^2$. The coefficient of this term is not fixed by symmetry and depends on the model. In the case under study it is approximately proportional to $\gamma\xi$. By using the second term appearing in \cref{phinnl} in \eqref{zzzNLNG} we obtain
\begin{multline}
		B_{\scriptscriptstyle\vp'^{\scriptscriptstyle2}}(k_1,k_2,k_3)=2\pi\xi\frac{\gamma}{H}\left(\frac{H}{\rho f}\frac{m^2}{f}\right)^4\nu_{\mathcal{O}}^2\int\dd\tau_1 \dd\tau_2\dd\tau_3
		G_{k_1}(0,\tau_1)G_{k_2}(0,\tau_2)G_{k_3}(0,\tau_3)\times\\
	\times \left[\partial_{\tau_1}G_{k_2}(\tau_1,\tau_2)\partial_{\tau_1}G_{k_3}(\tau_1,\tau_3)+{\rm 2\,\, perms.}\right]  .
\label{F2}
\end{multline}
For the corresponding amplitude in the equilateral configuration defined in \cref{fnldef} we get
\eq{
	f_{\rm NL}^{\text{eq}}=\frac{5\pi\xi}{3}\frac{\gamma}{H}\left(\frac{4^{\alpha-2} \Gamma (\alpha -1) \Gamma (\alpha )^2}{ \Gamma(\alpha-\frac{1}{2})\Gamma(2\alpha-\frac{1}{2})}\right)^{-2}\int_0^\infty\dd{y}\tilde{G}(0,y)\left(\int_y^\infty\dd{y'}\tilde{G}(0,y')\p_y\tilde{G}(y,y')\right)^2\,.
	\label{fNL2}
}
The numerical plot of \eqref{fNL2} as a function of $\gamma/H$ for fixed value of $\xi$  is shown in figure~\ref{fNLplot}. The important point is that unlike the previous case the asymptotic value is more or less independent of $\gamma/H$. In fact one can check numerically that the asymptotic value for $\gamma/H\gtrsim10$ is simply given by\footnote{More specifically, one can check that the value of the integral in \cref{fNL2} is $\approx0.04/(\gamma/H+0.75)(\gamma/H-0.59)$ for $\gamma/H\gtrsim10$.} 
\eq{
f_{\rm NL}^{\text{eq}}\simeq1.1\,\xi\,.
}
Therefore, for large enough values of the friction coefficient the contribution of $\vp'^2$ to the bispectrum becomes subdominant. 

Finally, the last contribution results from the operator $\vp'\delta\Oc_S$, for which we find
\begin{multline}
		B_{\scriptscriptstyle\vp'\delta\Oc_{\scriptscriptstyle S}}(k_1,k_2,k_3)=2\pi\xi\left(\frac{H}{\rho f}\frac{m^2}{f}\right)^4\nu_{\mathcal{O}}^2
		\int\dd{\tau}\dd{\tau'}\tau\bigg\{\Big[G_{k_2}(0,\tau)G_{k_3}(0,\tau')\partial_\tau G_{k_3}(\tau,\tau')\\+\{k_2\leftrightarrow k_3\}\Big] G_{k_1}(0,\tau) +{\rm 2\,\, perms.}\bigg\} .
\label{F3}
\end{multline}
The corresponding $f_{\rm NL}^{\text{eq}}$ is given by
\eq{
	f_{\rm NL}^{\text{eq}}=\frac{10\pi\xi}{3} \left(\frac{4^{\alpha-2} \Gamma (\alpha -1) \Gamma (\alpha )^2}{ \Gamma(\alpha-\frac{1}{2})\Gamma(2\alpha-\frac{1}{2})}\right)^{-2}\int_0^\infty\dd{y}y\,\tilde{G}(0,y)^2\int_y^\infty\dd{y'}\tilde{G}(0,y')\p_y\tilde{G}(y,y')\,.
	\label{fNL3}
}
Also in this case, as shown in figure~\ref{fNLplot}, the value of $f_{\rm NL}^{\text{eq}}$ approaches a constant for large $\gamma/H$. One can check numerically that for $\gamma/H\gtrsim10$ we have\footnote{More specifically, the integral in \cref{fNL3} can be approximated as $\approx-0.1/(\gamma/H-0.59)$ for $\gamma/H\gtrsim10$.}
\eq{
f_{\rm NL}^{\text{eq}}\simeq-5.7\xi\,.
}
Note that, as it is clear from \cref{infeqn2} and figure~\ref{fNLplot}, for most values of $\gamma/H$, the dominant contribution to $f_{\rm NL}^{\rm eq}$ comes from the $\vp'\delta\Oc_S$ term, unless $\gamma/H \gtrsim 20\xi$. Notice also that unlike the previous cases, this contribution does not vanish as $\gamma/H\to0$. The $\vp'\delta\Oc_S$ contribution depends on $\partial(\ln \nu_{\cal O})/\partial{\dot{\phi}_0} \simeq (4\pi \xi)/(\rho f)$ and as such it has an order-one uncertainty. Its size is about the same order of the present experimental uncertainty $\Delta f_{\rm NL}^{\text{eq}} \sim 50$: an improvement of the experimental bounds is thus very crucial for this scenario.
The plots of the shapes for $\vp'^2$ and $\vp'\delta\Oc_S$ are shown in figure~\ref{shapeplot}. 

\fg{
	\includegraphics[width=0.5\textwidth,trim={0 2cm 0 3cm},clip]{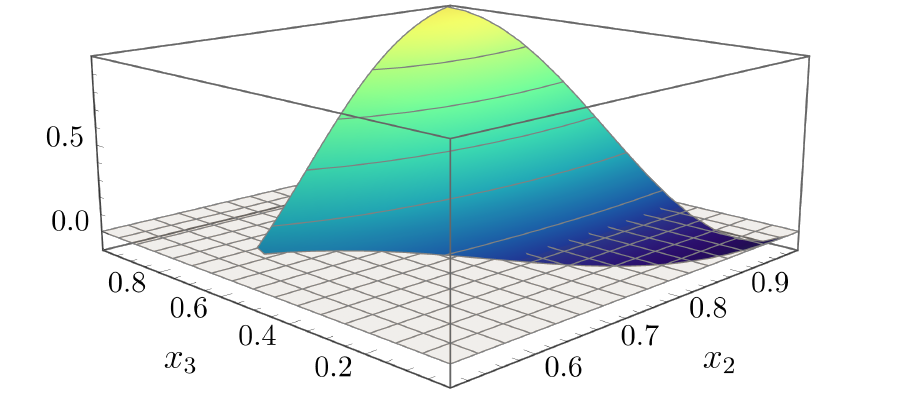}\hfil\includegraphics[width=0.47\textwidth,trim={0 2cm 0 2.5cm},clip]{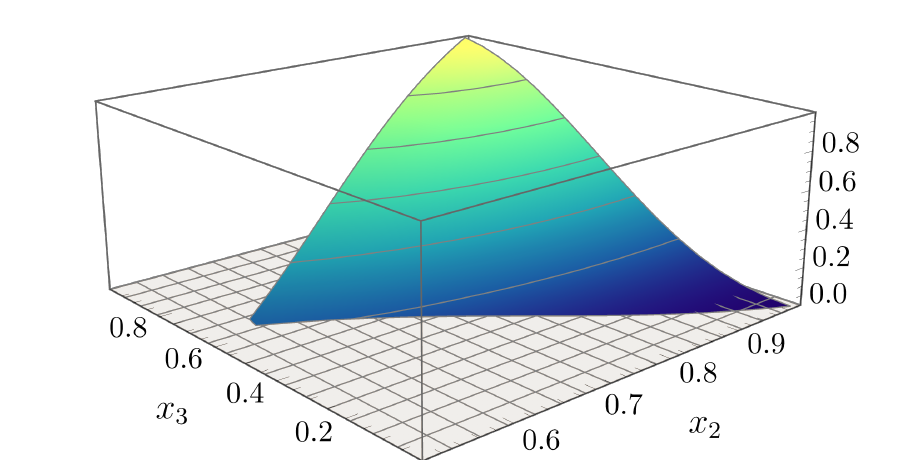}
	\caption{Shapes for the terms $\vp'^2$ (left) and $\vp'\delta\Oc_S$ (right).}
	\label{shapeplot}
}

\subsection{Non-Gaussian Noise}
\label{sec:intrinsicNG}
Another source of deviation from Gaussianity comes directly from the non-Gaussian statistics of the noise $\delta\Oc_S$. Using the leading-order solution \eqref{varphiinh}, obtained in linear response theory,  we can compute at late times:
\begin{multline}
	\langle\vp_{\vec k_1}\vp_{\vec k_2}\vp_{\vec k_3}\rangle=-\left(\frac{m^2}{f}\right)^3\int \dd{\tau}_1\dd{\tau}_2\dd{\tau}_3 a(\tau_1)^2a(\tau_2)^2a(\tau_3)^2G_{k_1}(0,\tau_1)G_{k_2}(0,\tau_2)G_{k_3}(0,\tau_3)\times
	\\
	\times \langle\delta\Oc_S(\tau_1,k_1)\delta\Oc_S(\tau_2,k_2)\delta\Oc_S(\tau_3,k_3)\rangle\,,
	\label{phiphiphiNG}
\end{multline}
which is the straightforward generalization of \eqref{varphivarphi} to the three-point correlator. 
Following the same logic as for \cref{noisek-2} for the two-point function of the noise, we get
\begin{equation}
	\langle\delta\Oc_S(\tau_1,k_1)\delta\Oc_S(\tau_2,k_2)\delta\Oc_S(\tau_3,k_3)\rangle
	\simeq {(2\pi)^3}
	\delta(\vec k_1+\vec k_2+\vec k_3)\delta(\tau_1-\tau_2)\delta(\tau_1-\tau_3) \, H^8 \tau_1^8 \nu_{\mathcal{O}^3} \, ,
	\label{noisek-3}
\end{equation}
which is manifestly local in time and space, and is valid as long as we restrict to long-wavelength perturbations by the locality approximation. The parameter $\nu_{\mathcal{O}^3}$ controls the amplitude of the non-Gaussianity of the noise and is fixed by the microscopic physics. We have worked out the details of the calculation of the correlation functions of $\delta\Oc_S$ in \cref{app:2pt}.
In particular, we find (see \cref{3ptOlocal})
\eq{
\nu_{\mathcal{O}^3}\simeq\frac{\e^{6\pi\xi}}{\pi^2m^2}\,.
\label{nuO3}
} 
Combining eqs.~\eqref{phiphiphiNG} and \eqref{nuO3}, and using the definition of the bispectrum in \cref{zeta3}, we obtain
\eq{
	B_{\scriptscriptstyle\Oc^3}(k_1,k_2,k_3)=\left(\frac{H}{\rho f}\frac{m^2}{f}\right)^3\nu_{\mathcal{O}^3}\int\frac{\dd{\tau}}{a(\tau)^2}G_{k_1}(0,\tau)G_{k_2}(0,\tau)G_{k_3}(0,\tau)\,.
}
By using \cref{fnldef}, we get
\eq{
	f_{\rm NL}^{\text{eq}}=\frac{5}{18}\frac{\int\dd{y}y^2\tilde{G}(0,y)^3}{\left(\int\dd{y}\tilde{G}(0,y)^2\right)^2}\frac{\nu_{\mathcal{O}^3}H^2}{\frac{H}{\rho f}\frac{m^2}{f}\nu_{\mathcal{O}}^2}\,,
	\label{fnlnoise1}
}
where we have expressed the power spectrum in terms of the integral over the Green's function as in \cref{phiphi2}. The last ratio in \cref{fnlnoise1} can be computed as
\eq{
\frac{\nu_{\mathcal{O}^3}H^2}{\frac{H}{\rho f}\frac{m^2}{f}\nu_{\mathcal{O}}^2}=\frac{16\pi\xi}{\gamma/H}\frac{m^2}{M^2}\,,
}
which implies 
\eq{
	f_{\rm NL}^{\text{eq}}\simeq\frac{40\pi}{9}\xi\frac{m^2}{M^2}\,,
	\label{fnlnoise}
}
where we have used a numerical estimate for the integrals in \cref{fnlnoise1}. Unlike in the previous cases discussed in~\cref{sec:nonlinearNG}, this contribution is typically expected to be smaller than unity (at most order-one if the separation between $m$ and $M$ is not too large), as a result of the hierarchy \eqref{hierarchy}.

%%%%%%%%%%%%%%%%%%%%%%%%%%%%%%%%%%%%%%%%%%%%%%%%%%%
\section{Conclusions and Future Directions}
In this work, we have introduced a model of dissipative inflation with scalar particle production. As opposed to previous implementations based on a coupling to gauge fields \cite{Anber:2009ua,Barnaby:2011vw}, in our model particle production takes place on parametrically sub-horizon scales. As we discussed, this has multiple advantages. First of all, a local approximation in the dynamics prevents the system from developing nonlocal responses and memory effects, which may induce unwanted large oscillations in the background solution. 
In addition, by virtue of the central limit theorem and the large occupation on short scales, the statistics of the perturbations is naturally close to Gaussian. This makes the predictions compatible with the observational constraints without the need for extra ingredients or assumptions. As a result, we have mostly been able to proceed analytically. 
We have obtained the expression \eqref{power} for the power spectrum as a function of the scales in the problem, and calculated the size of non-Gaussianity (\cref{sec:nongaussianity}) with the leading contribution to $f_{\text{NL}}^{\text{eq}}$ given by~\cref{fNL3}. Importantly, the size of the $f_{\text{NL}}^{\text{eq}}$ is  always large provided that cosmological perturbations are sourced by the ADOF. The size is indeed at reach of current and upcoming non-Gaussianity searches.
In the limit of very large friction, $(\gamma/H)\gtrsim 100$, the leading contribution would instead come from~\cref{fNLgammaH} which is determined by the nonlinearly realized Lorentz invariance.
The model \eqref{action}  is made robust by an approximate shift invariance on $\phi$ and a softly-broken global $U(1)$ symmetry of the ADOF, which protects the inflaton potential and the hierarchy \eqref{hierarchy} against large radiative corrections. In this sense, our model is, to the best of our knowledge, the first robust explicit example realizing the EFT of dissipative inflation of Ref.~\cite{LopezNacir:2011kk}.

Our results suggest a number of interesting future research directions. First of all, it would be interesting to study tensor modes in the model \eqref{action}.  It is well known that vacuum fluctuations of the graviton field during inflation might not be the only source contributing to the tensor power spectrum and B-mode polarization in the cosmic microwave background. 
Secondary gravitational waves can be produced as a result of an instability in the system, which converts a sizeable fraction of the inflaton's energy density into an extra sector coupled to it, which is, in turn, responsible for sourcing gravitational radiation.
From a qualitative viewpoint, as opposed to cases like axion inflation where the process occurs coherently over Hubble scales~\cite{Anber:2009ua,Sorbo:2011rz,Cook:2011hg,Barnaby:2011qe,Anber:2012du}, in our model graviton emission can happen locally through decay or scattering of the ADOF fields after they are produced on sub-horizon scales~\cite{Senatore:2011sp,Mirbabayi:2014jqa}. A precise estimate of the strength of gravitational waves in terms of the various scales appearing in \eqref{action} requires however a dedicated study and is left for future work.

In this paper, we mostly focused on the regime in which the $\chi$ production gives a sizeable backreaction, i.e.~friction, on the rolling inflaton background. In this regime, the exponential growth of the $\chi$ modes is automatically limited by the backreaction that slows down the inflaton background. The situation is different if one considers a regime of weak backreaction: in this case, the $\chi$ production is exponentially sensitive to the parameters of the background solution. Even a tiny time dependence of the background parameters will give a large deviation from scale invariance in the ADOF production. In analogy with the case of gauge-field instability, one is naturally led to consider a scenario that starts from a small backreaction at large scales and evolves to a large backreaction later on, as a consequence of the exponential sensitivity \cite{Barnaby:2011qe}. This can induce an enhancement in the scalar power spectrum on short scales and therefore be relevant for primordial black hole formation (for a review of the production mechanisms see \cite{Ozsoy:2023ryl}) and for gravitational-wave production on scales of interest for the future interferometers \cite{Cook:2011hg}. An important negative aspect of the scalar production we are discussing is the loss of chirality of the gravitational waves, which is a striking signature of the gauge field production. Conversely, one has the big advantage of avoiding the background instability recently found in these models \cite{Domcke:2020zez,Caravano:2022epk,Peloso:2022ovc,Figueroa:2023oxc}. 

In addition, it would be interesting to explore the possibility of thermalization in our model, with the produced ADOF particles eventually reaching a nonzero temperature phase. In such a `multi-fluid' scenario, where a hot plasma coexists with the inflaton, nontrivial effects may occur. The interaction with the plasma may alter the instability and the evolution of the system, potentially affecting reaching an equilibrium configuration.  
Even when an attractor at nonzero temperature can be reached, the thermalization process is expected anyway to be qualitatively different from heating by sphalerons in the non-abelian minimal warm inflation case~\cite{Berghaus:2019whh}, where the inflaton couples to a Yang--Mills gauge group. It would be interesting to see how this may affect for instance the shape and size of non-Gaussianity, and compare it with previous works in this context, e.g., \cite{Berera:1995ie,Berera:2008ar,Ferreira:2017lnd,Berghaus:2019whh,Mirbabayi:2022cbt}, as well as to study gravitational-wave production in the thermal phase in our model~\cite{Klose:2022rxh}.

On more general grounds, our work can also be thought of as a proof of concept that, as one may have expected, dissipative inflation is not an exclusive feature of couplings to gauge fields, but can be realized more in general. Aside from scalars, it will be worth in the future exploring the case of other spins, such as fermions. In the small-backreaction regime, the analysis of the power spectrum and  non-Gaussianity from fermion production during axion inflation has been done in Ref.~\cite{Adshead:2018oaa}.

Finally, although considering generic functions of $(\partial\phi)^2$, we included in \eqref{action} only quadratic operators in $\chi$. Self-interactions of the ADOF are on the other hand generically expected to be generated quantum mechanically through coupling to $\phi$ and one may wonder how they can affect the particle production. 
We leave this and the other open questions for future work.

\paragraph{Acknowledgements.} It is a pleasure to thank Mehrdad Mirbabayi for numerous insightful discussions. We also thank Marco Peloso, Leonardo Senatore, and Lorenzo Sorbo. S.K.~is supported in part by the NSF grant PHY-1915314 and the U.S.~DOE Contract DE-AC02-05CH11231. S.K.~thanks the Center for Cosmology and Particle Physics at New York University for hospitality while this work was in progress. B.S.~thanks IAP and APC in Paris for hospitality while this work was in progress. L.S.~is currently supported by the Centre National de la Recherche Scientifique (CNRS).

%%%%%%%%%%%%%%%%%%%%%%%%%%%%%%%%%%%%%%%%%%%%%%%%%%%
%%%%%%%%%%%%%%%%%%%%%%%%%%%%%%%%%%%%%%%%%%%%%%%%%%%
\appendix

%%%%%%%%%%%%%%%%%%%%%%%%%%%%%%%%%%%%%%%%%%%%%%%%%%%
\section{Canonical Quantization of ADOF}
\label{app:canquant}

In this section, we briefly review the canonical quantization procedure for the ADOF fields $\sigma_1$ and $\sigma_2$ in \eqref{eqschi1chi2dS2} neglecting inflaton perturbations. The Lagrangian for the ADOF is given by
\begin{equation}
\begin{split}
	\mathcal{L}=&-\frac{1}{2}(\p\s_1)^2-\frac{1}{2}(\p\s_2)^2+\frac{1}{2}\left(M^2+\frac{9H^2}{4}-m^2\right)\s_1^2+\frac{1}{2}\left(M^2+\frac{9H^2}{4}+m^2\right)\s_2^2
	\\
&+\rho(\s_1\dot\s_2-\s_2\dot\s_1)\,.
\label{Lsigma}
\end{split}
\end{equation}
Note that because of the coupling between the fields we cannot quantize them separately (see also \cite{Assassi:2013gxa}). The conjugate fields are
\eq{
\pi_1 \equiv \fdv{\mathcal{L}}{\dot\s_1}=\dot\s_1-\rho\s_2\,,\qquad \pi_2 \equiv\fdv{\mathcal{L}}{\dot\s_2}=\dot\s_2+\rho\s_1\,.
}
The standard commutation relations,
\begin{align}
	[\s_i(t,\bfx),\s_j(t,\bfx')]=0,~[\s_i(t,\bfx),\pi_j(t,\bfx')]=i\delta_{ij}\delta(\bfx-\bfx'),~[\pi_i(t,\bfx),\pi_j(t,\bfx')]=0, 
\end{align}
can be rewritten as,
\eq{
[\s_i(t,\bfx),\s_j(t,\bfx')]=0\,,~[\s_i(t,\bfx),\dot\s_j(t,\bfx')]=i\delta_{ij}\delta(\bfx-\bfx')\,,~[\dot\s_i(\bfx),\dot\s_j(\bfx')]=-2(\tau_2)_{ij}\rho\delta(\bfx-\bfx')\,,
\label{comsigma}
}
in which $\tau_2$ is the second Pauli matrix. Note that the last set of commutation relations are not the standard ones because of the last term in \cref{Lsigma}, which modifies the conjugate fields. 
One could remedy this by equivalently working with the rotated field basis
\eq{
\tilde\s_i(t,\bfx)=R_{ij}(t)\s_j(t,\bfx)\,,\qquad R=\begin{pmatrix}
	\cos(\rho t) & - \sin(\rho t) \\
	\sin(\rho t) &  \cos(\rho t) 
\end{pmatrix}\,.
\label{tildesigma}
}
The effect of the time-dependent orthogonal transformation \eqref{tildesigma} is to remove the derivative mixing in the Lagrangian \eqref{Lsigma}.
The drawback is however that the equations of motion for $\tilde\s_i$ involve oscillating functions of $\rho t$, which complicates an analytic treatment of the mode functions (see \cref{app:wkb}).
Therefore, we work with the unrotated $\sigma$ fields.

Following the standard treatment, we decompose the fields in terms of creation and annihilation operators as
\begin{equation}
\s_i(t,\bfx) = \int \frac{\D^3\bfk}{(2\pi)^{3}} \e^{i\bfk\cdot\bfx} \left[
(F_k(t))_{ij}\hat{a}_j(\bfk)  + (F_k^*(t))_{ij}\hat{a}^\dagger_j(-\bfk)  
\right] \, ,
\end{equation}
where we have packed the mode functions into a $2\times2$ matrix $F_k(t)$, which  depends on $k=|\bfk|$ and $t$, and where $\hat{a}_i(\bfk)$  and $\hat{a}_i^\dagger (\bfk)$ are annihilation and creation operators, respectively. Requiring the validity of the commutation relations \eqref{comsigma}, along with 
\begin{equation}
	[\hat{a}_i(\bfk), \hat{a}_j^\dagger(\bfk')] = {(2\pi)^3} \delta_{ij}\delta(\bfk-\bfk')\,,
\end{equation}
implies\footnote{Note that the right-hand side of \eqref{eqsFF3} is nonzero  because $\dot\s_i$ are not the canonical conjugate fields to $\sigma_i$. One would have found the right-hand side of \eqref{eqsFF3} to vanish in the rotated basis $\tilde{\sigma}_i$, defined in \cref{tildesigma}.}
\begin{subequations}
	\label{eqsFF}
	\begin{align}
		F_{k}{ F}_{k}^{*T}- F^*_{k}{ F}_{k}^{T}  & = 0 \, ,
		\label{eqsFF2}\\
		F_{k}\dot{F}_{k}^{*T}- F^*_{k}\dot{F}_{k}^{T} & = i\mathbf{1} \, ,
		\label{eqsFF1}\\ 
		\dot{ F}_{k}\dot{ F}_{k}^{*T}- \dot{ F}^*_{k}\dot{ F}_{k}^{T} & =-2\rho\tau_2 \, .
		\label{eqsFF3}
	\end{align}
\end{subequations}
The mode functions satisfy the same equations of motion as the fields,
\begin{equation}
\label{eomFk}
\ddot{F}_k + 
 \begin{pmatrix}
 0 &-2\rho\\
 2\rho & 0
 \end{pmatrix} \cdot
\dot{F}_k + 
\begin{pmatrix}
 \frac{k^2}{a^2} - M^2  - \frac{9}{4}H^2+m^2 &0\\
 0 & \frac{k^2}{a^2} - M^2  - \frac{9}{4}H^2-m^2
 \end{pmatrix} \cdot  F_k
  =0  \, .
\end{equation}
Note that \eqref{eqsFF} and \eqref{eomFk} are not all independent.  For instance, taking the time derivative of \eqref{eqsFF1}, and using first eq.~\eqref{eqsFF3} and then the equation of motion \eqref{eomFk}, one recovers the condition \eqref{eqsFF2}.

The initial conditions are fixed at very early times when $k/a$ is very large. 
One can then neglect the mixing terms in \eqref{eomFk} and treat the fields $\s_i$  as just standard plane waves in Minkowski space.  
Imposing the Bunch--Davies initial conditions amounts to choosing the positive-frequency solution only for each $\s_i$. In terms of the mode function $F_k$ that solves \eqref{eomFk}, we require
\begin{equation}
	{ F}_k(t\rightarrow-\infty) \rightarrow  \frac{\e^{-i k \tau}}{\sqrt{2k/a}}\begin{pmatrix}
		1 & 0 \\ 0 & 1
	\end{pmatrix} \, ,
	\label{initialcondF}
\end{equation}
where $\tau$ is the conformal time, $\D \tau = \D t /a(t)$.\footnote{The factor of $a^{1/2}$ in \eqref{initialcondF} may look unfamiliar. It arises because we used cosmic time, instead of conformal time, to write the canonically normalized  Lagrangian for $\sigma_1$ and $\sigma_2$. }

%%%%%%%%%%%%%%%%%%%%%%%%%%%%%%%%%%%%%%%%%%%%%%%%%%%
\section{WKB Solution for the Mode Function Matrix $F$}
\label{app:wkb}
In this appendix, we derive the WKB solution of \cref{eomFk}. The final result is also summarized in \cref{fullWKB-main} in the main text. Note that each column of the matrix $F_k$ solves the same equation \eqref{eomFk}. Therefore, to obtain the general form of the solution, we can just replace $F_k$ in \cref{eomFk} with a 2-vector, which we will call $\vec F_{\rm{column}}$, and solve for it. The full solution for $F_k$ can be constructed afterward by replacing each column of $F_k$ with $\vec F_{\rm{column}}$, and imposing the boundary conditions~\eqref{initialcondF}.

We can find the  zeroth-order eikonal approximation for  $\vec F_{\rm{column}}$ by assuming that the coefficients of the equation \eqref{eomFk}  depend weakly on $t$ and that all the time dependence in the vector $\vec F_{\rm{column}}$ comes from a common phase factor (see e.g.~\cite{Weinberg:PhysRev.126.1899}), i.e.,
\eq{
	\vec F_{\rm{column}}=\vec Q(t)\exp(-i\int\dd{t}\,\omega(t))\,.
	\label{Fanzts}
}
This essentially amounts to neglecting the expansion of the Universe at the leading order.
In this limit, \cref{eomFk} becomes a set of coupled ordinary differential equations with constant coefficients. Plugging \eqref{Fanzts} in \cref{eomFk}, and neglecting derivatives of  $\omega$ and ${\vec Q}$, at the leading order we obtain the following algebraic system of equations:
\eq{
	D(\omega)\cdot\vec{Q}\equiv\begin{pmatrix}
		-\omega^2+\frac{k^2}{a^2}-M^2-\frac{9H^2}{4}+m^2&2i\rho\,\omega\\
		-2i\rho\,\omega&-\omega^2+\frac{k^2}{a^2}-M^2-\frac{9H^2}{4}-m^2
	\end{pmatrix}\cdot\vec{Q}=0\,.
	\label{wkbeq0}
}
This can be satisfied nontrivially only if  the determinant of  the matrix $D$ is zero, which  yields the following values for $\omega$ (see also eqs.~\eqref{omegapm} and \eqref{nu2} in the main text):
\eq{
	\omega_{\pm}^2=\left(\sqrt{\frac{k^2}{a^2}+\mu^2}\pm\rho\right)^2-\frac{m^4}{4\rho^2}\,,\qquad \mu^2\equiv\rho^2-M^2-\frac{9H^2}{4}+\frac{m^4}{4\rho^2}\,.
	\label{omegapmapp}
}
The eigenvectors associated with the zero eigenvalues $\omega_\pm$ of $D$ are then given by, 
\eq{
	\vec{Q}_{\pm}\propto\begin{pmatrix}
		&2i\rho\omega_{\pm}\\
		&\omega_{\pm}^2-\frac{k^2}{a^2}+M^2+\frac{9H^2}{4}-m^2
	\end{pmatrix}\,.
	\label{Qnull}
} 
In \eqref{Qnull} there is an overall time-dependent prefactor, which we did not write explicitly. This can be fixed by considering the next-to-leading order (NLO) of the WKB expansion, which requires keeping the time derivatives of $\omega$ and $\vec Q$. To this end,  let us first redefine for convenience the amplitude in \cref{Fanzts} as $\vec{Q}_{\rm full}$ and expand it as follows,
\eq{
	\vec{Q}_{\rm full}=\vec{Q}_\pm+\vec{Q}_{\rm NLO}+\dots\,,
}
where $\vec{Q}_\pm$ are the leading solutions \eqref{Qnull}, while $\vec{Q}_{\rm NLO}$ captures subleading terms of the same order of $\dot{\vec{Q}}_\pm$ and $\dot \omega$.
Expanding \cref{eomFk}  to NLO  yields
\eq{
	D(\omega_{\pm})\cdot \vec{Q}_{\rm NLO}-i\dot{\omega}_{\pm}\vec{Q}_{\pm}-2i\begin{pmatrix}
		\omega_{\pm}&-i\rho\\
		i\rho&\omega_{\pm}
	\end{pmatrix}\dot{\vec{Q}}_{\pm}=0\,,
\label{app:eqnlo}
}
where the matrix $D$  is the same as in \cref{wkbeq0}. 
Let us now multiply the equation \eqref{app:eqnlo} from the left by $\vec{Q}_{\pm}^\dagger$.  Since $D(\omega_\pm)$ is Hermitian and  solves $D(\omega_\pm)\cdot \vec{Q}_\pm=0$, the first term in \eqref{app:eqnlo} becomes zero, while the remaining terms can be rearranged as follows,
\eq{
	\dv{{}}{t}\left[\vec{Q}_{\pm}^\dagger\begin{pmatrix}
		\omega_{\pm}&-i\rho\\
		i\rho&\omega_{\pm}
	\end{pmatrix}\vec{Q}_{\pm}\right]=0\,.
}
Combining this with \cref{Qnull}, we can now solve for the time-dependent prefactor of $\vec Q_\pm$ and find
\begin{equation}
	\label{Qpmwkb-app}
	\vec Q_\pm  \equiv \frac{1}{2\sqrt{2}\left(\frac{k^2}{a^2}+\mu^2\right)^{1/4}}  \begin{pmatrix}
		\,u_{\pm}
		\\
		\mp\frac{i}{u_{\pm}}
	\end{pmatrix} \, ,
	\qquad\quad
	u_{\pm}\equiv\left[\frac{\omega_{\pm}}{\sqrt{\frac{k^2}{a^2}+\mu^2}\pm\left(\rho-\frac{m^2}{2\rho}\right)}\right]^{1/2} \,.
\end{equation}
Note that there is still an undetermined, overall constant factor in $\vec Q_\pm$ in \eqref{Qpmwkb-app}, which will be fixed by the boundary condition \eqref{initialcondF}.

So far, with the ansatz \eqref{Fanzts},  we have only taken into account the positive-frequency solutions with $\exp(-i\int\omega)$. Everything we have said above is also valid for the negative-frequency solutions with $\exp(+i\int\omega)$ with an appropriate complex conjugation. 
Thus, the final WKB result for $F_k$, to next-to-leading order,  can ultimately be written as
\eq{
	F_k= \sum_{\lambda=\pm} \Bigg(
	c^{(1)}_{\lambda}\vec{Q}_{\lambda}\,\e^{-\,i\int\omega_\lambda}
	+ \, d^{(1)}_{\lambda}\vec{Q}^*_{\lambda}\,\e^{+\,i\int\omega_\lambda} 
	\,\, ,\,\,  
	c^{(2)}_{\lambda}\vec{Q}_{\lambda}\,\e^{-\,i\int\omega_\lambda}
	+ \, d^{(2)}_{\lambda}\vec{Q}^*_{\lambda}\,\e^{+\,i\int\omega_\lambda} 
	\Bigg)\,,
	\label{F-full}
}  
where we have summed over the $\omega_\pm$ modes, as well as the positive and negative-frequency solutions. $c^{(i)}$ and $d^{(i)}$, where the superscript $i$ labels the columns of $F_k$, are constant coefficients. 

\subsection{Validity of the WKB Approximation} 
\label{app:WKBvalidity}
The WKB approximation is valid as long as  $\vert\dot\omega/\omega^2\vert\ll1$. 
In this section,   we  check this condition for the $\omega_-$ mode.\footnote{We will not discuss the  $\omega_+$ mode explicitly since it is straightforward to check, following analogous considerations, that its WKB solution is valid at all times during the evolution.} 
Recall that we are interested in the case in which $\omega_-^2$ becomes negative, triggering an instability in the ADOF sector, over a finite duration of time for each $k$-mode (see, e.g.,  figure~\ref{omega2plot}).
We shall separate the analysis into three different steps, reflecting the three regions of figure~\ref{omega2plot}: early times ($t<t_1$), intermediate times ($t_1<t<t_2$), and late times ($t>t_2$). 

Let us start from $t<t_1$. Using the expression for $\omega_{-}$ given in \cref{omegapmapp} one can see that, at very early times,
\eq{
	\Big\vert\frac{\dot\omega_{-}}{\omega^2_{-}}\Big\vert\sim\frac{aH}{k} \lesssim \frac{H}{M} \ll1\,,
} 
where we used that, for $t<t_1$, $k/a\gtrsim M$ (see \cref{ineqka}) and that $M\gg H$ (see the discussion in \cref{sec:localapprox} and, in particular, \cref{instwindowc}).
At late times, after the second turning point ($t>t_2$), we find instead
\eq{
	\Big\vert\frac{\dot\omega_{-}}{\omega^2_{-}}\Big\vert\sim\frac{Hk^2}{a^2M^3} \lesssim \frac{H}{M}  \ll1\,,
} 
where we assumed that $\rho\sim M$ and used that $\omega_-\sim M$ for $k/a\lesssim M$ (see \cref{sec:localapprox} for more details about the scales).
Inside the instability region ($t_1<t<t_2$), the ratio can be  roughly estimated as
\eq{
	\Big\vert\frac{\dot\omega_{-}}{\omega^2_{-}}\Big\vert\sim\frac{HM^2}{m^2\rho}\sim\frac{1}{8\xi}\frac{m^2}{\rho M}\ll1\,,
}  
where we introduced $\xi\simeq m^4/(8H\rho M^2)$ (see \cref{app:xi} below and the conditions \eqref{hierarchy}).
Clearly, these estimates are valid as long as $t$ is sufficiently far from $t_1$ and $t_2$. Close to the turning points, $\omega_-^2 \simeq 0$ and the ratio $\vert\dot\omega_-/\omega_-^2\vert$ formally diverges, signalling a breakdown of the WKB approximation.  How far from the turning points $t$ has to be in order for the WKB expansion to remain valid can be easily estimated by Taylor expanding $\omega^2_-\simeq [\D\omega_-^2/\D t]_{t=t_j}(t-t_j)$, for $j=1,2$, and using  \eqref{omegapmapp}. The final result is that $\vert\dot\omega_-/\omega_-^2\vert\ll1$ provided that $H \vert t-t_j \vert \gg (H/m)^{2/3}$. As a consistency check, note that the lower bound for $\vert t-t_j \vert$ is much less than a Hubble time, and is also small compared to the width $\Delta t$ of the instability band: $ (H/m)^{2/3}/(H\Delta t)\sim (M^3H/m^4)^{2/3}\sim 1/(4 \, \xi^{2/3})\ll1$, where we used \cref{instwindownarrow} and the expression \eqref{app:xi} for $\xi$ with $\rho\sim M$.

As long as $H \vert t-t_j \vert \gg (H/m)^{2/3}$, we can thus apply \eqref{F-full}. The problem is hence reduced to finding a matching procedure at the turning points that allows us to determine the unknown coefficients $c^{(i)}$ and $d^{(i)}$ in \eqref{F-full}. This will be the content of the next subsections. The final result is summarized in \cref{fullWKB-main} in the main text. A comparison between the WKB result \eqref{fullWKB-main} and the numerical solution of the equation \eqref{eomFk} is also displayed in figure~\ref{wkbnumeric}. Notice that the agreement is always very good except close to the edges of the grey bands, corresponding to the turning points $t_1$ and $t_2$, where the WKB solution (dashed lines in figure~\ref{wkbnumeric}) breaks down, as discussed above.
\fg{
	\centering
	\includegraphics[width=0.9\textwidth,trim={2cm 2.5cm 2cm 2.5cm},clip]{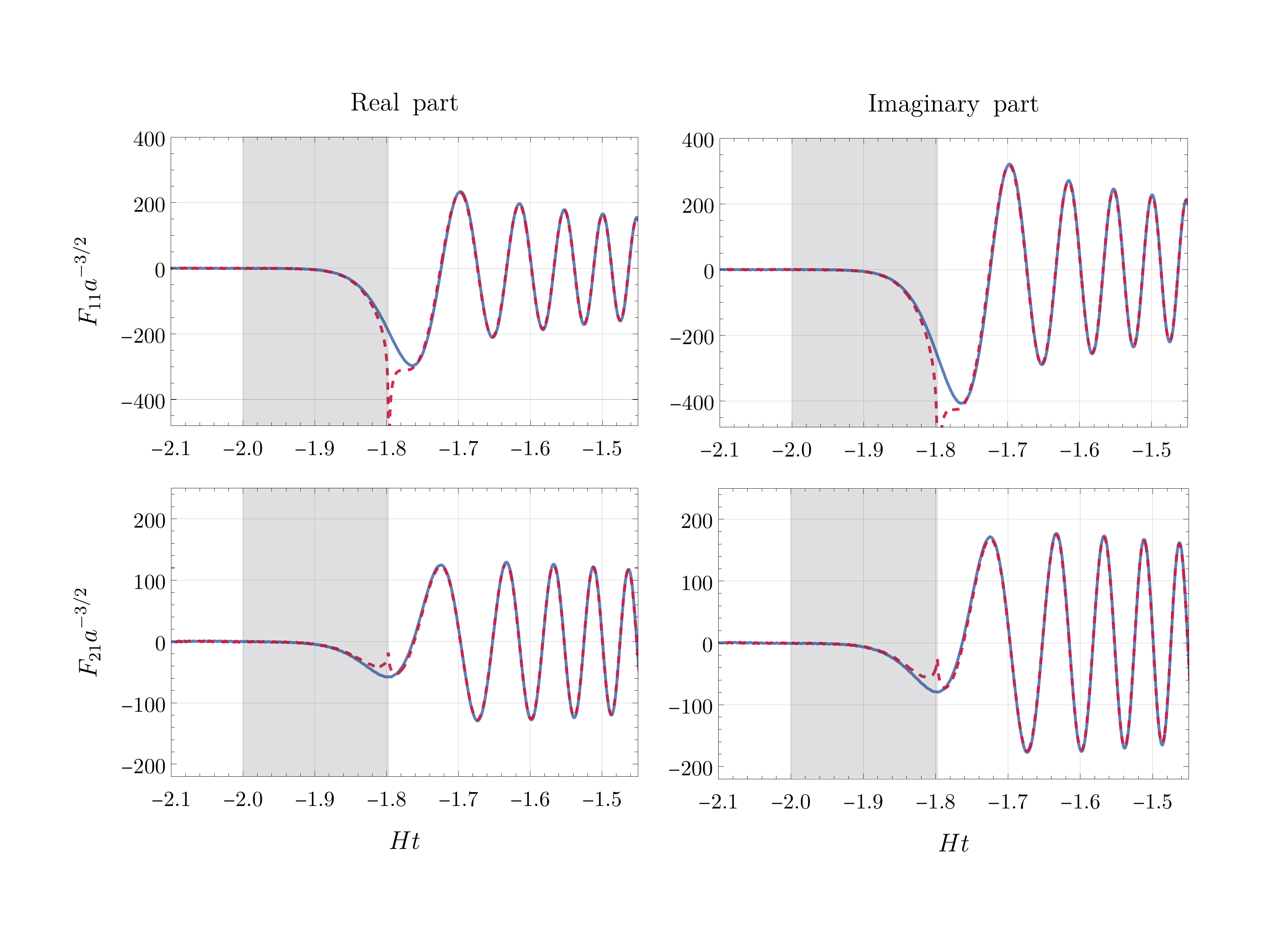}
	\caption{We show the real and imaginary parts of the entries in the first column of the matrix $F_k$ that solves  \eqref{eomFk}. In particular, we compare the WKB solution (dashed lines) in \cref{fullWKB-main} with the numerical solution (solid lines), given the initial conditions \eqref{initialcondF}.  The grey bands correspond to the instability region, where $\omega_-^2\leq 0$.}
	\label{wkbnumeric}
}

\subsection{Matching Conditions at the Turning Points and Growth Factor}

\paragraph{Early-time WKB solution.}
Starting from the general solution \eqref{F-full} and imposing the Bunch--Davies initial condition \eqref{initialcondF} yields the following result for $F_k(t)$ at times $t<t_1$:
\eq{
	F_k=\bigg(\vec{Q}_+\,\,,\,\, i\vec{Q}_+\bigg)\e^{-i\int^t\omega_+}+\bigg(\vec{Q}_-\,\,,\,\,-i\vec{Q}_-\bigg)\e^{-i\int^t\omega_-}\,.
	\label{Fin-app}
}

\paragraph{Matching at the first turning point.}
For intermediate times $t$ in the interval $t_1<t<t_2$,  $\omega_-$ is purely imaginary---recall that $\omega_+$ instead remains real throughout the whole evolution. 
Neglecting the exponentially decreasing solution and the oscillatory $\omega_+$ modes---which do not get amplified---the  form of the WKB solution inside the instability band, from \cref{F-full}, is
\eq{
	F_k=\bigg(b_1\vec{Q}_-\,\,,\,\, b_2\vec{Q}_-\bigg)\e^{+\int_{t_1}^t|\omega_-|}\,,
	\label{Fmid-app}
} 
where $b_1$ and $b_2$ are constant coefficients, which we  shall fix by matching with \eqref{Fin-app} at $t=t_1$. In \eqref{Fmid-app}, the expression for the vector $\vec{Q}_-$ can be read off from \cref{Qpmwkb-app}, where the frequency $\omega_-$ should be understood as $\omega_- \mapsto i \vert\omega_-\vert$.
To find $b_1$ and $b_2$, we will follow the procedure of Refs.~\cite{Landau:1991wop,Dufaux:2006ee}, which consists of performing an analytic continuation of the solution in complex $t-$plane. Starting from the solution at $t<t_1$, instead of matching with the solution valid at $t>t_1$ by approaching the turning point $t_1$ along the real axis, we shall instead walk around it by following a semicircle in the complex plane, as in figure~\ref{complt}. The radius of the contour has to be sufficiently large such that the WKB expansion remains valid (see the subsection~\ref{app:WKBvalidity} above for more quantitative considerations), but small enough so that we can expand
\begin{equation}
	\omega_-\approx\kappa_1(t_1-t)^{1/2}\,,\qquad
	\vec{Q}_-\approx\frac{1}{2\sqrt{2\rho+m^2/\rho}}
	\begin{pmatrix}
		0  \\ \frac{i m/\sqrt{\rho\kappa_1}}{(t_1-t)^{1/4}}
	\end{pmatrix}\,,  \qquad \text{for }\,\vert t-t_1\vert \rightarrow 0 \, ,
	\label{WKBQ2b}
\end{equation}
\fg{
	\centering
	\includegraphics[width=0.6\textwidth,trim={0 1cm 0 5cm},clip]{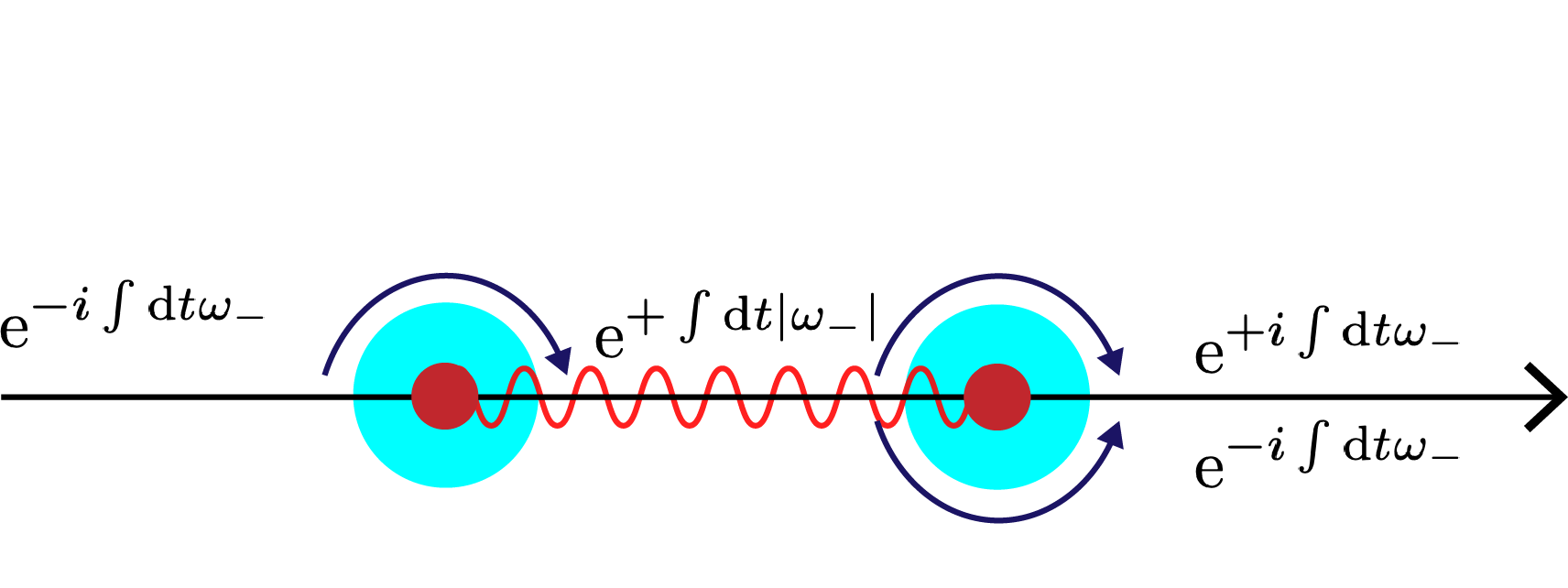}
	\caption{Analytic continuation and contours in the complex $t-$plane. The red circles denote the location of the turning points. In the cyan region, the WKB approximation breaks down, and analytic continuation, along with matching, is done avoiding that region.}
	\label{complt}
}%
where we have defined $\kappa_1=\sqrt{\frac{2Hm^2(M^2+m^2)}{2\rho^2+m^2}}>0$.\footnote{{Eq.~\eqref{WKBQ2b} is valid both when $t$ approaches $t_1$ both from above $(t\rightarrow t_1^+)$ and when it approaches it from below $(t\rightarrow t_1^-)$. When $t\rightarrow t_1^+$, the denominator in the second entry of $\vec{Q}_-$ should be interpreted as $(t_1-t)^{1/4}= \e^{i \frac{\pi}{4}}\vert t_1-t\vert^{1/4}$. Similar considerations will apply to \cref{WKBQ2b-second} below for the second turning point.}}
Comparing the analytic continuation of the solution \eqref{Fin-app} along the semicircle in  figure~\ref{complt} with \cref{Fmid-app}, we get  the matching conditions
\eq{
	b_1=i b_2 =\e^{-i\theta_1}\,,
\label{Bmatch}
} 
where $\theta_1$ is the accumulated phase up to the first turning point, $\theta_1\equiv \int^{t_1}\dd{t'}\omega_{-}(t')$. Plugging \eqref{Bmatch} into \eqref{Fmid-app} yields the expression in the second line of \eqref{fullWKB-main}.

Once inside the instability region $t_1<t<t_2$, the mode \eqref{Fmid-app} starts growing exponentially. The final relative amplification at $t=t_2$ can be easily computed as
\eq{
	\pi\xi\equiv\int_{t_1}^{t_2}\dd{t'}|\omega_{-}(t')|=\int_{\sqrt{M^2+m^2}}^{\sqrt{M^2-m^2}}\frac{\dd{\kp}}{H\kp}|\omega_{-}|\,,
\qquad \kp\equiv \frac{k}{a} \, ,
}  
which, using \cref{omegapmapp}, admits a closed-form solution:
\begin{equation}
	\xi= \frac{1}{H}\left[\rho-\sqrt{\rho^2-\frac{M^2+\frac{9H^2}{4}-\sqrt{(M^2+\frac{9H^2}{4})^2-m^4}}{2}}\right] \, .
	\label{xi-app}
\end{equation}
This expression can be simplified using the hierarchy \eqref{hierarchy} in \cref{sec:localapprox}, from which we obtain
\begin{equation}
	\xi \simeq \frac{m^4}{8H \rho M^2}\,.
\label{app:xi}
\end{equation} 

\paragraph{Matching at the second turning point.}
Similarly to the previous case, the expression for $F_k$ at times $t>t_2$ can be read off  from \cref{F-full}, where we shall neglect the  $\omega_+$ modes, which are exponentially suppressed compared to the $\omega_-$ modes. We can thus write, for $t>t_2$,
\eq{
	F_k=\bigg(c_1 \vec{Q}_-\,\,,\,\,c_2\vec{Q}_-\bigg)\e^{-i\int_{t_2}^t\omega_-}+\bigg(d_1\vec{Q}^*_-\,\,,\,\, d_2 \vec{Q}^*_-\bigg)\e^{+i\int_{t_2}^t\omega_-}\,,
	\label{Faf-app}
}
where $c_1$, $c_2$, $d_1$ and $d_2$ are constant coefficients. To fix them, we shall proceed as before by analytically continuing the solution \eqref{Fmid-app} and matching with \eqref{Faf-app} across $t=t_2$. The only difference is  that, in order to reproduce the positive and negative-frequency components of \eqref{Faf-app}, we now have to analytically continue  \eqref{Fmid-app} along two different semicircles, one in the upper half and one in the lower half of the complex $t-$plane  \cite{Landau:1991wop,Dufaux:2006ee}---see figure~\ref{complt}.  
Let us first Taylor expand around $t_2$,
\begin{equation}
	\omega_-\approx\kappa_2(t-t_2)^{1/2}\,,\qquad
	\vec{Q}_-\approx\frac{1}{2\sqrt{2\rho-m^2/\rho}}
	\begin{pmatrix}
		\frac{i m/\sqrt{\rho\kappa_2}}{(t-t_2)^{1/4}}
		\\
		0
	\end{pmatrix}\,,  \qquad \text{for }\,\vert t- t_2\vert \rightarrow 0 \, ,
	\label{WKBQ2b-second}
\end{equation}  
where we have now defined $\kappa_2=\sqrt{\frac{2Hm^2(M^2-m^2)}{2\rho^2-m^2}}>0$. 
Then, going  along the upper contour in figure~\ref{complt}, one gets the  $\exp(+i\int\omega_{-})$ solution, which allows to  fix $d_1$ and $d_2$  in terms of $b_1$. Following  the lower contour allows instead to find  $c_1$ and $c_2$.\footnote{Note that, in the latter case, one has to cross a branch cut, which is responsible for the relative phase between $d_j$ and $c_j$ ($j=1,2$) in \cref{app:match2}.}
The final result of the matching is
\eq{
	c_1=ic_2=\e^{\pi\xi}b_1\,,\qquad d_1=id_2=i\e^{\pi\xi}b_1\,.
\label{app:match2}
}

\section{Background Expectation Value of ${\cal O}$}
\label{app:signofO}
In this section, we provide more details on the estimate \eqref{eqschiphibis-last2} for the expectation value of the operator ${\cal O} \equiv -i (\chi^2-\chi^{*2})$ on the background. In particular, we will check the sign of the last term in \cref{scalareqback} and show that it is consistent with the sign of the standard Hubble friction  $H\dot{\phi}_0$.

From \cref{eqschiphibis-last}, we can write
\begin{equation}
\label{eqschiphibis-lastapp}
-\frac{im^2}{f} \ev{\chi^2-\chi^{*2}}  =
\frac{m^2}{2\pi^2f a^3}\int_0^\infty \D k \, k^2 [F_k(t)\cdot F_k^\dagger(t)]_{12} \, .
\end{equation}
Using the WKB solution \eqref{fullWKB-main} with \eqref{Qpmwkb-app}, and restricting the integral in \eqref{eqschiphibis-lastapp} over the instability band---which provides the leading contribution to \eqref{eqschiphibis-lastapp}---we have:
\begin{equation}
F_k (t_1< t<t_2)  =
	 \frac{\e^{-i\theta_1}}{2\sqrt{2}\left(\frac{k^2}{a^2}+\mu^2\right)^{1/4}}  \begin{pmatrix}
	 u_- & -i u_-
	 \\
	 \frac{i}{u_-} &  \frac{1}{u_-}
	 \end{pmatrix}
	 \e^{\int_{t_1}^t|\omega_-|  \D t } \, ,
\end{equation}
\begin{equation}
\label{eqschiphibis-lastapp2}
-\frac{im^2}{f} \ev{\chi^2-\chi^{*2}}  \simeq
\frac{m^2}{8\pi^2f}\int_{M^2-m^2}^{M^2+m^2} \frac{\D (\frac{k}{a}) \, \frac{k^2}{a^2}}{\left(\frac{k^2}{a^2}+\mu^2\right)^{1/2}} \frac{-i u_-^2}{\vert u_-\vert^2} \e^{2\int_{t_1}^t|\omega_-|  \D t } \, .
\end{equation}
Now, notice that, from \cref{omegapmapp},
\begin{equation}
	\omega_{-}^2=\left(\sqrt{\frac{k^2}{a^2}+\mu^2}-\rho-\frac{m^4}{4\rho^2}\right)  \left(\sqrt{\frac{k^2}{a^2}+\mu^2}-\rho+\frac{m^4}{4\rho^2}\right)  <0
\end{equation}
for $t_1<t<t_2$. Thus, it follows in particular that 
\begin{equation}
-i u_{-}^2 = \frac{ \vert \omega_{-}\vert}{\sqrt{\frac{k^2}{a^2}+\mu^2}-\rho+\frac{m^2}{2\rho}} >0 \, .
\end{equation}
As a result, \eqref{eqschiphibis-lastapp} is a finite integral of a positive function, therefore $-\frac{im^2}{f} \ev{\chi^2-\chi^{*2}} >0$.

%%%%%%%%%%%%%%%%%%%%%%%%%%%%%%%%%%%%%%%%%%%%%%%%%%%%%%%%%
\section{Statistics of the Noise}
\label{app:2pt}
In this section we study the correlation functions of the stochastic part of the operators that appear in our model. 
All the operators that are quadratic in the ADOF fields in our model can be cast in the compact form
\eq{
	\Oc=\frac{1}{a^3}A_{ij}\s_i\s_j\,,
	\label{O}
} 
where the matrix $A$ defines the operator and we are implicitly  summing over repeated indices.
For instance, for $\Oc=-i(\chi^2-\chi^*{}^2)$, the matrix $A$ is
\begin{equation}
A=\begin{pmatrix}
	0&1\\1&0
\end{pmatrix}.
\label{Amtr}
\end{equation}
 The expectation value of the operator is then given by the expression
\eq{
	\ev{\Oc}=\frac{1}{a^3}\int\frac{\dd[3]{\bfk}}{(2\pi)^3}\Tr(A^TF_kF^\dagger_k)\,,
}
where $F_k$ is the mode functions matrix given in \cref{fullWKB-main} (see \cref{app:wkb} for more details). The stochastic part of the operator is defined as the deviation from its expectation value, i.e.,~$\delta\Oc_S\equiv[\Oc-\ev{\Oc}]_{\varphi=0}$. As discussed in the main text, we are interested in $\Oc=-i(\chi^2-\chi^*{}^2)$. Therefore, in the following we will concentrate on the case \eqref{Amtr} and we will  only briefly comment on the behavior of the other operators.

\subsection*{Two-point Function}

We are interested in computing  the two-point function in Fourier space. After some straightforward  algebra, we obtain
\eq{
	\langle\delta\Oc_S(t,\bfk)\delta\Oc_S(t',\bfk')\rangle=\frac{(2\pi)^3\delta(\bfk+\bfk')}{a(t)^3a(t')^3}\int\frac{\dd[3]{\bfp}}{(2\pi)^3}\Tr[A^T\F_p(t,t')(A+A^T)\F^T_{|\bfk-\bfp\,|}(t,t')]\,,
	\label{2ptstoc-gen}
}
where we introduced
\eq{
	\F_{p}(t,t')\equiv F_p(t)F_p^\dagger(t')\,.
	\label{Fcal}
}
We get an exponential enhancement if both modes with momenta $\vec p$ and $\bfk-\bfp$ have gone through the instability band or are just about to leave it. 
Therefore, in order to evaluate $\F$ from \cref{Fcal} we can  use the late-time WKB solution for $F$ given in last line of \cref{fullWKB-main}. One can check that the contribution from the modes inside the instability band only changes the final result by an $O(1)$ amount. By direct calculation one obtains 
\eq{
	\F_{p}(t,t')=\frac{\e^{2\pi\xi}}{4\big[(\frac{p^2}{a^2}+\mu^2)(\frac{p^2}{a'^2}+\mu^2)\big]^{1/4}}
	\begin{pmatrix}
		|u_-u'_-|(EE'^*+iEE')&|\frac{u_-}{u'_-}|(iEE'^*+EE')\\
		|\frac{u'_-}{u_-}|(-iEE'^*+EE')&\frac{1}{|u_-u'_-|}(EE'^*-iEE')
	\end{pmatrix}+ {\rm c.c.}\,.
	\label{Ftild}
}
In the above expression we have introduced the oscillating phase
\eq{
	E\equiv \e^{-i\int_{t_2(p)}^{t}\omega_{-}(\tilde t)\dd{\tilde t}}\,,
	\label{E}
}
where $t_2(p)$ is defined as the time at which the mode with momentum $p$ exits the instability band. 
In addition, the expression for $u_-$ can be read off from \cref{Qpmwkb-app}. 
Note that the prime symbol in \eqref{Ftild} is just a shorthand, meaning that in the argument of the function $t$ is replaced by $t'$, and it should not be confused with a derivative with respect to conformal time. 
Note also that in \eqref{Ftild} we have neglected the exponentially subdominant mode associated with the $\omega_+$ solution.

In the expression~\eqref{2ptstoc-gen} we have two factors of $\F$ computed at two different scales, associated with $p$ and $q\equiv |\bfk-\bfp|$. 
However, we will eventually be interested in the correlation functions of the noise at large separations of points, i.e., for $k$ much smaller than $p$ and $q$.
Therefore, in the following, we shall neglect $k$ and just set $q\simeq p$. 

For the operator of interest $\Oc=-i(\chi^2-\chi^*{}^2)$, the expression \eqref{2ptstoc-gen}  reduces to 
\begin{align}
	\langle\delta\Oc_S(t,\bfk)\delta\Oc_S(t',\bfk')\rangle &=\frac{4(2\pi)^3\delta(\bfk+\bfk')}{a(t)^3a(t')^3}\int\frac{\dd[3]{\bfp}}{(2\pi)^3}\left[(\F_p)_{11}(\F_{|\bfk-\bfp|})_{22}+(\F_p)_{12}(\F_{|\bfk-\bfp|})_{21}\right](t,t')  \nonumber \\
	&\simeq(2\pi)^3\delta(\bfk+\bfk')\frac{2\e^{4\pi\xi}}{a(t)^3a(t')^3}\int\frac{\dd[3]{\bfp}}{(2\pi)^3}\frac{\cos(2\int^{t}_{t_2(p)}\omega_p)\cos(2\int^{t'}_{t_2(p)}\omega_p)}{\big[(\frac{p^2}{a(t)^2}+\mu^2)(\frac{p^2}{a(t')^2}+\mu^2)\big]^{1/2}},
\label{eqc8}
\end{align}
where in the second line we have used \cref{Ftild} and  the approximation $q\simeq p$. 
Using simple trigonometric identities, the oscillating factors in the integrand can be recast in the form
\eq{
\frac{1}{2}\left[\cos(2\int^{t}_{t'}\omega_p)+\cos(2\int^{t}_{t'}\omega_p+4\int^{t'}_{t_2(p)}\omega_p)\right]\,.
}
Both terms are oscillating for generic values of $t$ and $t'$. This implies in particular that  the integral in \eqref{eqc8} averages the oscillations to zero and becomes negligible when the difference $\Delta t\equiv t-t'$ is large.
Instead, for small  separations $\Delta t$, one gets
\eq{
	\langle\delta\Oc_S(t,\bfk)\delta\Oc_S(t',\bfk')\rangle '\simeq\frac{\e^{4\pi\xi}}{4\pi^2a^3}\int^{\sim M}\dd{x}\frac{x^2}{x^2+\mu^2}\bigg(\e^{-2i\omega(x)\Delta t}+\,\rm{c.c.}\bigg)\,,
} 
where we have defined $x\equiv p/a$ and where the upper limit of integration is given the position of the instability band $\sqrt{M^2-m^2}\sim M$. Since in this range $\omega(x)$ is a decreasing function of $x$ (until it becomes zero at $x=\sqrt{M^2-m^2}$) the integral is dominated by its upper limit. Expanding  around the upper limit,
\eq{
	\omega(x)\approx\left(\frac{2m^2\sqrt{M^2-m^2}}{2\rho^2-m^2}\right)^{1/2}\sqrt{\sqrt{M^2-m^2}-x}\,.
} 
Then, the result of the integral can be approximated as
\eq{
	\int^{\sim M}\dd{x}\frac{x^2}{x^2+\mu^2}\bigg(\e^{-2i\omega(x)\Delta t}+ \, {\rm c.c.} \bigg)\approx \frac{M}{m}\frac{\sin(2m\Delta t)}{\Delta t}\,.
}
We are interested in cases $\Delta t \sim 1/H$   so that $m\Delta t\gg1$.
In this limit we can use the formal identity $\frac{\sin(x/\epsilon)}{x}\to\pi\delta(x)$, which holds in the limit  $\epsilon\to0$.\footnote{The more precise statement is that $\lim_{\epsilon\to0}\int\dd{x}\frac{\sin(x/\epsilon)}{\pi x}f(x)=f(0)$ for any smooth function $f$. Therefore in this limit, we can treat the sinc function as a delta function.
} Finally we get
\eq{
	\langle\delta\Oc_S(t,\bfk)\delta\Oc_S(t',\bfk')\rangle\simeq(2\pi)^3\delta(\bfk+\bfk')\frac{\delta(t-t')}{a^3}\left(\frac{\e^{4\pi\xi}}{4\pi^2}\frac{M}{m}\right)\,.
}

\subsection*{Three-point Function}
In this section, we derive the three-point function of the noise, which we used in \cref{sec:intrinsicNG} to estimate the  non-Gaussianity induced on $\varphi$ within linear-response theory. 
Similarly to the evaluation of power spectrum, after some straightforward algebra, we get 
\begin{multline}
	\langle\delta\Oc_S(t,\bfk)\delta\Oc_S(t',\bfk')\delta\Oc_S(t'',\bfk'')\rangle'=
	\\
	= \frac{8}{a(t)^3a(t')^3a(t'')^3}\int\frac{\dd[3]{\bfp}}{(2\pi)^3}\Tr[A\F_p(t,t')A\F_{|\vec k'+\vec p|}(t',t'')A\F_{|\vec k-\vec p|}^T(t,t'')]\,,
	\label{3ptstoc-gen}
\end{multline}
where we have assumed that $A$ is a symmetric matrix.
Once again we are interested in $\Oc=-i(\chi^2-\chi^*{}^2)$. 
By analogous considerations as in the case of power spectrum, we can neglect $k$, $k'$ with respect to $p$. 
First, let us look at the modes  after the instability band. 
Using the WKB solution \eqref{fullWKB-main} we obtain for the integrand of \cref{3ptstoc-gen}:
\eq{
\e^{6\pi\xi}\frac{\cos(2\int^{t}_{t_2(p)}\omega_p)\cos(2\int^{t'}_{t_2(p)}\omega_p)\cos(2\int^{t''}_{t_2(p)}\omega_p)}{\big[(\frac{p^2}{a(t)^2}+\mu^2)(\frac{p^2}{a(t')^2}+\mu^2)(\frac{p^2}{a(t'')^2}+\mu^2)\big]^{1/2}}\,.
\label{3ptstoc}
}
As a result, it is easy to see that for large separation among the three times $t$, $t'$ and $t''$ the integrand is highly oscillating and makes therefore the integral \eqref{3ptstoc-gen} negligible. 
Hence, we shall set $t=t'=t''$  to evaluate the leading contribution to the integral. The most relevant modes are the ones that just came out of the instability band and are slowly varying at $t$. Thus, each of the cosines can be approximated as
\eq{
	\cos\left[m \sqrt{\frac{M-x}{M}}(t-t_{2}(p))\right] = \cos\left[\frac{m}{H} \sqrt{\frac{M-x}{M}}\log\frac{M}{x}\right] \sim \cos\left[\frac{m}{H} \left({\frac{M-x}{M}}\right)^{3/2}\right] \, ,
}
where in the first step we rewrote $t-t_{2}(p)$ in terms of the comoving momentum $x = p/a$, while in the second one we made an approximation valid close to the instability band $x \sim M$. One can now do the integral over momenta and get
\eq{
	\ev{\delta\Oc_S^3}_{ t=t'=t''}'\simeq\frac{4\e^{6\pi\xi}}{a^6\pi^2}\int^{\sim M}\dd{x}\frac{x^2}{(x^2+\mu^2)^{3/2}}\cos^3\left(\frac{m}{H}\left(\frac{M-x}{M}\right)^{3/2}\right)\simeq\frac{\e^{6\pi\xi}}{a^6\pi^2}\left(\frac{H}{m}\right)^{2/3}\,.
	\label{3ptstoc-app}
}
This gives the correlation of the noise at equal times; in order to write it at different times one needs to estimate how fast the correlation decays when say $t$ and $t'$ are separated. The integral in $x$ above gets contribution for $(M-x)/M \lesssim (H/m)^{2/3}$. In this range the frequency of oscillations of the modes is smaller than $m (H/m)^{1/3}$: one starts seeing cancellations when the difference in time $t-t'$ becomes of order $m^{-1} (H/m)^{-1/3}$. This gives the width of the delta functions and 
we can approximate the correlation of the noise as
\eq{
	\langle\delta\Oc_S(t,\bfk)\delta\Oc_S(t',\bfk')\delta\Oc_S(t'',\bfk'')\rangle\simeq(2\pi)^3\delta(\vec k+\vec k'+\vec k'')\frac{\delta(t-t')\delta(t-t'')}{a^6}\frac{\e^{6\pi\xi}}{\pi^2}\frac{1}{m^2}\,.
	\label{3ptOlocal}
}

\bibliographystyle{utphys}
\addcontentsline{toc}{section}{References}
\bibliography{biblio}

\end{document}